\documentclass[traditabstract]{aa}
\usepackage{graphicx}
\usepackage{txfonts}
\usepackage{longtable}
\usepackage[authoryear]{natbib}
\bibpunct{(}{)}{;}{a}{}{,}

\def\ltsima{$\; \buildrel < \over \sim \;$}
\def\simlt{\lower.5ex\hbox{\ltsima}}
\def\gtsima{$\; \buildrel > \over \sim \;$}
\def\simgt{\lower.5ex\hbox{\gtsima}}
\newcommand{\kms}{km~s$^{-1}$}

\begin{document}
   \title{Na-O anticorrelation and HB. IX. Kinematics of the program clusters\thanks{Based on observations collected at ESO telescopes under programs 072.D-507, 073.D-0211 and 083.D-0208.}}
   \subtitle{A link between systemic rotation and HB morphology?}

   \author{M. Bellazzini\inst{1}, A. Bragaglia\inst{1}, E. Carretta\inst{1}, R.G. Gratton\inst{2}, 
   S. Lucatello\inst{2}, G. Catanzaro\inst{3}, F. Leone\inst{4}}
         
      \offprints{M. Bellazzini}

   \institute{INAF - Osservatorio Astronomico di Bologna,
              Via Ranzani 1, 40127 Bologna, Italy\\
            \email{michele.bellazzini@oabo.inaf.it} 
            \and
            INAF - Osservatorio Astronomico di Padova,
            Vicolo dell'Osservatorio 5, 35122 Padova, Italy
            \and
            INAF - Osservatorio Astronomico di Catania, 
            Via S.Sofia 78, 95123 Catania, Italy
            \and
            Dipartimento di Fisica e Astronomia, Universitˆ di Catania, 
            Via S.Sofia 78, 95123 Catania, Italy
            }

     \authorrunning{M. Bellazzini et al.}
   \titlerunning{Na-O anticorrelation and HB. IX. Kinematics of the program clusters.}

   \date{Accepted for publication on A\&A }

\abstract{We use accurate radial velocities for 1981 member stars in 20 Galactic globular clusters, collected within our large survey aimed at analyzing the Na-O anti-correlation, to study the internal kinematics of the clusters. We performed the first systematic exploration of the possible connections between cluster kinematics and the multiple populations phenomenon in GCs. We did not find any significant correlation between Na abundance and either velocity dispersion or systemic rotation.
We searched for systemic rotation in the eight clusters of our sample that lack this analysis from previous  works in the literature (NGC2808, NGC5904, NGC6171, NGC6254, NGC6397, NGC6388, NGC6441, and NGC6838). These clusters are found to span a wide range of rotational amplitudes from $\sim$0.0~\kms~(NGC6397) to $\sim$13.0~\kms~(NGC6441). We found a significant correlation between the ratio of rotational velocity to central velocity dispersion ($V_{rot}/\sigma_0$) and the horizontal branch morphology parameter (B-R)/(B+R+V). The ratio $V_{rot}/\sigma_0$ is found to correlate also with metallicity, possibly hinting at a significant role for dissipation in the process of formation of globular clusters; $V_{rot}$ is found to correlate well with (B-R)/(B+R+V), $M_V$, $\sigma_0$, and [Fe/H]. All these correlations strongly suggest that systemic rotation may be intimately linked with the processes that led to the formation of globular clusters and the stellar populations they host.}


\maketitle
%

\section{Introduction}
\label{intro}

Our ideas on Galactic globular clusters (GC) have been revolutionized in the last few years by the discovery of multiple sequences in their color magnitude diagrams \citep[CMD; see][and references therein]{piotto, milone} and by evidence that the well-known star-to-star spread in the abundance of light elements (e.g. O, Na, Mg, Al) is not due to some mixing mechanism, but was imprinted in the gas from which cluster stars formed \citep[see][for discussion and references]{rc01,g01,g_araa}. While we lack a fully satisfying model for GC formation accounting for the variety of observational phenomena that are being unveiled, the notions that most GCs formed more than one generation of stars during the first few $10^8$~yr of life and that the subsequent generations were - at least partially - polluted by chemical elements produced by previous generations is now generally accepted \citep[see, e.g.][for a detailed discussion]{2808,Dec07,Dec10,VDA08,annib,alvio,IQR,VC11,bekkimod,shae}.

In the previous papers of this series, the chemical signatures of this self-enrichment process were studied in detail, with special attention to the anti-correlation between [Na/Fe] and [O/Fe], using a large sample ($>2000$) of stars in 19 GCs, for which medium-to-high resolution spectra were obtained with the Giraffe and UVES fiber-fed spectrographs, mounted at VLT@ESO \citep[see][for details, results and discussion]{p1n2808,p5n6441,p7gir,p8uve}. The main results of this large project have been considered in a wider context and correlated with other characteristics of the clusters, like the horizontal branch (HB) morphology, in \citet{scala,IQR}, and \citet{2nd3rd}. In the present contribution we use an obvious byproduct of the spectroscopic survey, i.e. relatively large sets of very accurate radial velocities (RV), to explore the kinematics of the 19 program clusters \citep[][Pap-VII hereafter]{p7gir}, plus NGC1851 from \citet[][see Table~\ref{Tab_sam}, below]{n1851}.

These samples, which were collected for a very different scientific goal, are far from ideal to provide a full and detailed characterization of the cluster kinematics. The typical number of stars per cluster ($\la 150$) is clearly not comparable to the most recent dedicated studies, based on several hundred RVs 
\citep[see e.g.,][]{lane1,lane2,lane3,sollima}, in some cases coupled with proper motions \citep{vandeven,bosch,mc47}. Furthermore, the radial distribution of the stars does not cover all the relevant radial ranges, because of limitations imposed by the fiber allocation process, and also because to obtain the most reliable chemical abundances, it is actually preferable to avoid the most crowded regions of the clusters where contamination from stars other than the intended target can become an issue. In particular most of the stars in our samples lie beyond the half-light radius ($r_h$) of the clusters (see Table~\ref{Tab_sam}, below). On the other hand, the membership to the parent clusters of all the stars considered in this study has been established not only on the basis of radial velocity cuts but also on very detailed chemical analysis of each individual star\footnote{In the particularly challenging case of NGC6441, which lies in the dense field of the Galactic bulge, distance from the cluster center was also taken into account when establishing membership \citep{p5n6441}.}. Moreover, our samples provide the unprecedented possibility of a systematic search for correlations between the chemical composition and the kinematics of the cluster stars.

The plan of the paper is the following. In Sect.~\ref{sample} we briefly describe the global sample and assess the quality of the RV estimates. In Sect.~\ref{kinchem} we explore the possible connection between the kinematics and the chemical abundance of Na, which is taken as the most reliable tracer of the light elements' self-enrichment at the basis of the Na-O anticorrelation. In Section~\ref{disp} we consider in more detail the RV distribution of the clusters whose kinematics have been previously studied only by means of integrated spectra of their center and/or with samples of RV for individual stars smaller than ours. In addition we search for systemic rotation in all the program clusters for which this analysis has not been performed in previous studies. In Sect.~\ref{summ} we briefly summarize and discuss the main results of the analysis, including some intriguing correlations between rotation and other cluster parameters, which are found here for the first time.

\section{The sample}
\label{sample}

\begin{figure}
   \centering
   \includegraphics[width=\columnwidth]{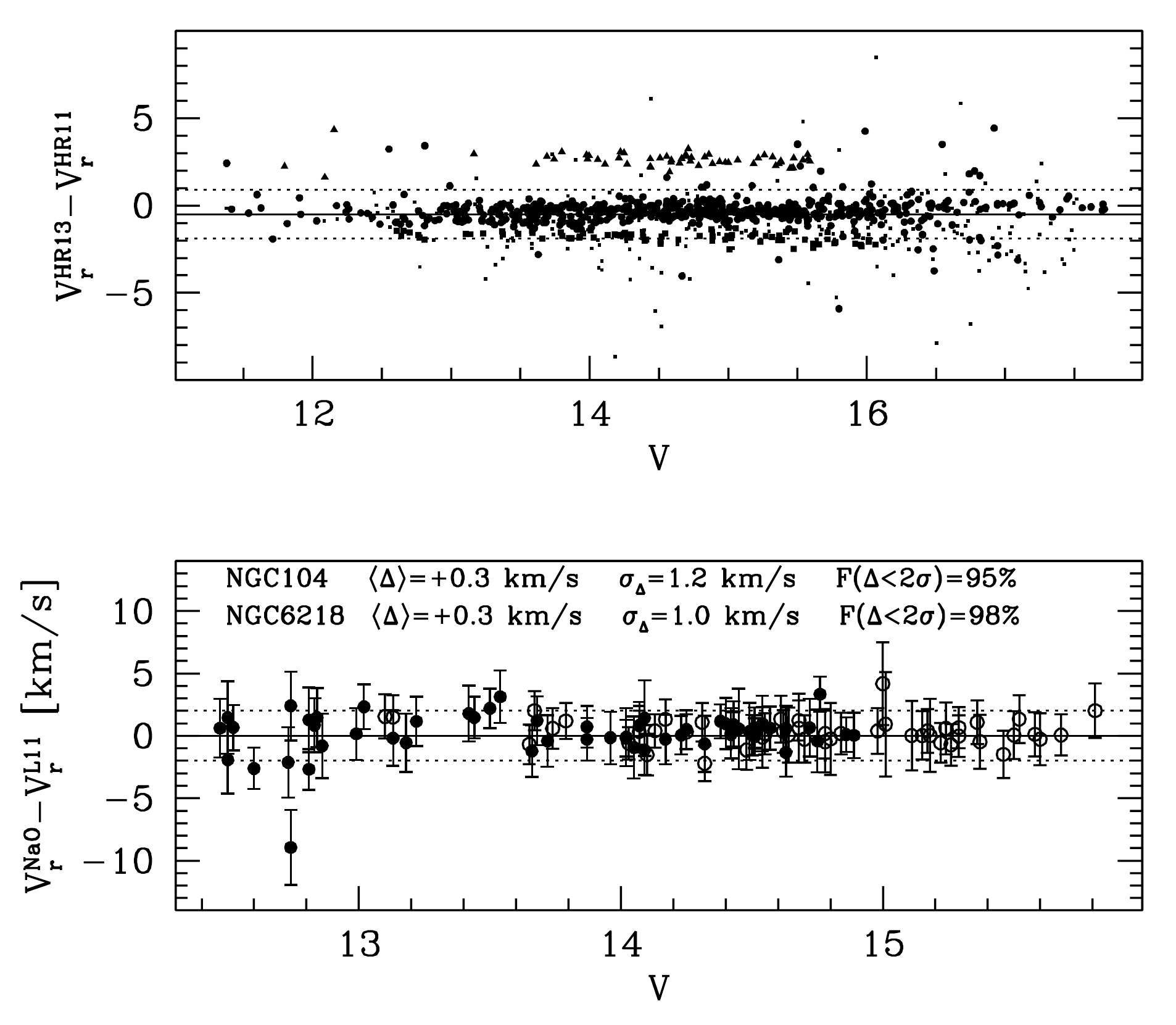}
     \caption{{\em Upper panel:} comparison between RVs estimated from spectra obtained with the HR13 and HR11 grisms. Different symbols correspond to different clusters. The continuous line is drawn at a typical value for the mean difference ($\Delta V_r=-0.5$~\kms), and the dotted lines enclose the $\pm 2\sigma$ range about the mean,
where a typical value of $\sigma=0.7$~\kms ~has been adopted (see Table~\ref{Tab_sam}).
{\em Lower panel:} comparison with the external dataset by L11 for two representative clusters, NGC104 (filled circles) and NGC6218 (empty circles). The continuous line marks $\Delta V_r=0.0$~\kms, the dotted lines enclose the $\pm 2\sigma$ range. The fraction of stars lying within $\pm 2\sigma$ is also reported
[$F(\Delta<2\sigma)$]. } 
        \label{compest}
    \end{figure}

In the following we use the symbols $V_r$ and RV interchangeably  to indicate heliocentric line-of-sight velocities. We make use of the data collected in the \citet[][H96 hereafter]{h96} catalog: in all cases we refer to the 2010 version of this catalog, publicly available at {\tt www.physics.mcmaster.ca/Globular.html}. The coordinates of the cluster centers are taken from \citet{ng06}, when available, and from H96 in the other cases. The parameters of the \citet[][K66]{k66} or \citet[][W75]{w75} models best-fitting the surface brightness profiles of the clusters are taken from \citet[][MvM05]{mcv}, specifically the core and tidal radii $r_c$, $r_t$, while half-light radii $r_h$, lacking in MvM05,  are from H96.
We adopt the metallicity scale by \citet{scala}. \citet{kbin} has recently demonstrated that in relatively dense and massive clusters like those considered here, the effects of binary stars on the global kinematics should be minor or negligible (e.g. leading to an increase in the observed velocity dispersion $\la 0.5$~\kms, even for binary fractions as high as 100 percent, see their Fig.~10). Since, in addition, GCs typically have binary fractions $\la 20$ percent \citep{antobin,milbin}, we consider binaries as a negligible factor for our analysis\footnote{The very low number of stars showing RV differences from repeated measures larger than $2\sigma$ in the lower panel of Fig.~\ref{compest}, below, also supports this conclusion.}.

Any detail about the measure of RV and chemical 
abundances, as well as target selection and membership can be found in the previous papers of the series
\citep{p1n2808,p5n6441,p7gir,p8uve,n1851}. All the data used in this paper are publicly available in electronic form at the CDS\footnote{\tt http://cdsweb.u-strasbg.fr/} (see the original papers).

The program clusters are listed in Table~\ref{Tab_sam}. 
The second column of the table reports the number of  member stars having a valid measure of RV. The number of stars per sample ranges from 30 to 156 stars per cluster, with a mean of 99.
Each of these stars also has at least a measure of iron abundance, and in many cases the abundance of other elements is also available (Na, in particular). The selected stars not only have RV that is fully compatible with the systemic velocity of the cluster but also display chemical abundance pattern identifying them as cluster members. The fraction of sample stars lying within the half-light radius of the cluster is listed in the third column of the table.

In the large majority of clusters most stars have been observed using two FLAMES-GIRAFFE set-ups, the high-resolution gratings HR11 (centered on 5728 \AA) and HR13 (centered on 6273 \AA), which were respectively chosen to measure the Na doublets at 5682-5688 \AA ~and 6154-6160 \AA, and the [O I] forbidden lines at 6300, 6363 \AA, as well as several lines of Fe-peak and $\alpha$-elements. The spectral resolutions are R = 24200 (for HR11) and R = 22500 (for HR13), at the center of the spectra. 
The fifth column of the table reports the number of stars having valid RV measures from spectra obtained with both set-ups. These stars therefore have two independent RV estimates and can be used to estimate the internal consistency and accuracy of our radial velocities. This is especially useful since the pipeline used for the survey does not provide  uncertainties on the individual RV estimates. However, it has been demonstrated by comparison with other independent sets of RV estimates that the adopted pipeline provides RV estimates with typical errors that are smaller than 1~\kms ~from spectra obtained with the same instrumental set-up \citep{m54all}\footnote{See also \citet{panci07}, for the accuracy of RV estimates obtained with FLAMES/GIRAFFE HR13 spectra by comparison with independent datasets.}.

The comparison between the RV estimates obtained from the two GIRAFFE set-ups is shown in the upper panel
of Fig.~\ref{compest}. The mean difference and the standard deviation of the difference between the two sets of estimates are reported in the 4th and 5th columns of Table~\ref{Tab_sam}. It can be appreciated that both the mean difference and the r.m.s. are lower than 1~\kms ~in the large majority of cases. The small differences in the zero-point between the two RV sets are not a reason for concern in the present analysis.
We transformed HR11 RVs into the HR13 system by applying the shifts listed in the 4th column of Table~\ref{Tab_sam}, then we obtained a single RV estimate per star by adopting $RV_{HR13}$ as reference value, and the corrected $RV_{HR11}$ value when the HR13 velocity is missing. Velocities from HR11 were chosen as a reference since the associated spectra typically have higher signal-to-noise ratios than HR13 ones.
On the other hand, we have to note that the HR13 - HR11 comparison reveals that three clusters have RV estimates of significantly lower accuracy than the rest of the sample. These are NGC4590, NGC6397, and NGC7078, three among the most metal-poor clusters of the sample. The lower quality of the RV measures for these clusters are likely accounted for by the extreme weakness of the spectral lines due to the extremely low metal content, possibly coupled with the relatively high temperature of the target stars. The latter effect depends on the actual distribution in color and magnitude of the targets, which can differ from cluster to cluster. For instance, considering the case of two very metal-poor clusters having RV measures of different quality, the median $T_{eff}$ of the NGC6809 sample ($\sigma\Delta=0.4$~\kms) is 4797~K, while it is 5030~K for NGC7078 ($\sigma\Delta=4.6$~\kms). Cooler atmospheres imply more detectable lines, keeping all the other parameters fixed, thus allowing more accurate RV estimates.

\begin{table*}
  \begin{center}
  \caption{The sample and its internal RV accuracy}\label{Tab_sam}
  \begin{tabular}{lccccccl}
    \hline
    Clus.&[Fe/H] & $N_{tot}^a$ & F($r\le r_h)^b$&$\langle \Delta RV_{HR13-HR11}\rangle$& $\sigma_{\Delta}$& $N_{HR13-HR11}^c$&Notes$^d$\\
     &     &             &                &  [\kms]                                &   [\kms]           &\\
\hline
NGC104 & -0.76&147 & 0.37&-0.38 & 0.24 & 84  & L10a$^e$  \\									  
NGC288 & -1.32&110 & 0.39&-0.19 & 0.52 & 50  & L10b									      \\
NGC1851& -1.16& 84 & 0.00&  --- &  --- & --  & INT(I76,D97), 184 stars from \citet{sca11}   \\
NGC1904& -1.58& 58 & 0.09&-0.81 & 1.55 & 40  & INT(D97), 146 stars from \citet{sca11} 	      \\
NGC2808& -1.18&123 & 0.06&-0.43 & 0.54 & 63  & INT(I76,Z92)								      \\
NGC3201& -1.51&149 & 0.48&-0.09 & 0.87 & 56  & 420 stars from \citet{cote}						      \\
NGC4590& -2.27&121 & 0.23&-1.21 & 2.70 & 98$^*$ & L09 									      \\
NGC5904& -1.33&136 & 0.13&-1.78 & 0.30 & 86  & INT(Dub97), 46 stars from \citet{rs91} 	      \\
NGC6121& -1.18&103 & 0.67&-0.81 & 0.60 & 63  & L10b						      \\
NGC6171& -1.03& 33 & 0.00&-0.07 & 0.30 & 27  & 71 stars assembled from various sources by PM93$^f$  \\
NGC6218& -1.33& 79 & 0.02& 2.61 & 0.38 & 56  & L10b									      \\
NGC6254& -1.57&147 & 0.32&-0.36 & 0.63 & 61  & 25 stars from \citet{rs91}						      \\
NGC6388& -0.45& 36 & 0.00&-0.07 & 0.55 & 23  & INT(I76)								      \\
NGC6397& -1.99&144 & 0.35&-1.30 & 1.59 & 73$^*$ & INT(D97), 127 stars from \citet{MM91}		      \\
NGC6441& -0.44& 30 & 0.00&  --- &  --- & --  & INT(I76,D97)								      \\
NGC6752& -1.55&137 & 0.29&-0.42 & 0.46 & 69  & L10b									      \\
NGC6809& -1.93&156 & 0.42&-0.25 & 0.40 & 72  & L10a									      \\
NGC6838& -0.82& 39 & 0.49& 0.19 & 0.15 & 28  & $>100$ stars assembled from various sources by PM93\\
NGC7078& -2.33& 84 & 0.13&-0.26 & 4.57 & 58$^*$ & 1773 stars form \citet{bosch}\\
NGC7099& -2.33& 65 & 0.20& 0.10 & 1.63 & 30  & L09								      \\
\hline
\end{tabular}
\tablefoot{
$^a$ Total number of stars with a valid RV measure.\\
$^b$ Fraction of sample stars lying within the cluster half-light radius $r_h$.\\
$^c$ Number of stars having RV measure both from HR11 and HR13.\\
$^d$ Reference to the most recent and/or largest sample of radial velocities available for the cluster, and
used to study the cluster kinematics \citep[PM93=][]{pm93}.
INT indicates the presence of an estimate of the central velocity dispersion from integrated spectra; the associated reference is reported within brackets [I76 = \citet{I76}; P89 = \citet{P89}; Z92 = \citet{zag}; D97 = \citet{D97}]. The INT references are reported only when large samples of individual velocities are lacking. NGC6171 and NGC6254 lack INT estimates.
$^e$ Clusters included in the L09, L10a and L10b papers have been studied using samples of RV estimates
several hundreds individual stars per cluster. We refer to that series of papers for further references.\\
$^f$ Preliminary results from a sample of 107 stars have been presented by \citet{sca04}.\\
$^*$ Clusters with lower quality RV estimates (see text).
} 
\end{center}
\end{table*}

In the lower panel of Fig.~\ref{compest}, we show the comparison between the RV from the present study and those from the public dataset by \citet[][L11, hereafter]{lanecat} for two clusters in common between the two datasets taken as representative of the whole range of magnitude covered by our targets, i.e. NGC104 and NGC6752. The zero-points agree within less than 0.5~\kms ~and, most important, the standard deviation in the RV difference is $\le 1.2$~\kms, implying typical errors $\sim 1.2/\sqrt(2)\simeq 0.8$~\kms. We note that 95\%-98\% of the differences lie within $\pm 2.0\sigma$ from the mean, which is fully consistent with the expectation for a Gaussian distribution. These results are in excellent agreement with those presented in \citet{m54all}. 
In conclusion, internal and external comparisons consistently indicate that individual RV estimates in our sample have typical uncertainties smaller than 1~\kms, so their quality is fully adequate for studying the internal kinematics of GCs. It should be kept in mind that in the cases of NGC4590, NGC6397, and NGC7078 the uncertainties are larger, typically 2~\kms, or even slightly larger in the case of NGC7078.

The last column of Table~\ref{Tab_sam} lists the main sources of kinematic data for each cluster available in the literature, i.e. from the largest samples and/or the most recent and comprehensive studies. Eight clusters are shared with the survey by \citet[][hereafter L09, L10a, L10b, respectively]{lane1,lane2,lane3}, whose data have been published in L11. These papers report comprehensive kinematical analysis based on very large samples (several hundred member stars per cluster) of $R= 10000$ spectra (in the CaT region) obtained with the AAOmega instrument at the Anglo Australian Telescope, and can be considered as the state-of-the-art, at least from the observational point of view (see L11, for discussion). A few other clusters have been studied with smaller samples than in L11 but still larger than ours: \citet{sca11}, \citet{cote}, \citet{MM91}, or assembled by \citet[][PM93 hereafter]{pm93}. PM93 also provide comprehensive references for studies performed before 1993-1994, and it remains the main source for central velocity dispersions ($\sigma_0$) of Galactic GCs. In several cases estimates of $\sigma_0$ derived from integrated spectra of the cluster center \citep[from][]{I76,zag,D97} constitute a relevant complement to samples of RV for individual stars and, in some case, the only available measure of the cluster kinematics (NGC2808, NGC6388, NGC6441). NGC7078 has been the object of a very detailed and refined analysis based on nearly two thousand RV and proper motions by \citet{bosch}.

\section{Kinematics and chemical composition}
\label{kinchem}

There are only very few massive Galactic GCs whose stars display a spread in iron (and/or calcium) abundance, either large ($\omega$~Cen, see 
\citealt{jp10,mari_omega} and references therein; Terzan5, \citealt{ter5,liv_t5}) or small (M54, \citealt{mic54,m54om}; M22, see \citealt{gary,mari_m22} and references therein; NGC2419, \citealt{judy2419,iba2419,ju24b}; NGC1851, \citealt{n1851}). At present there is no firmly established case for a correlation between iron abundance (or metallicity, in general) and star kinematics in these clusters. Interesting correlations have been reported by various authors for $\omega$~Cen
\citep{norris,fpm,antosig,gra11}, but were not confirmed by other studies \citep{panci07,bellini_pm}, hence they remain to be further verified \citep[see also][]{bekkirot}. \citet{n1851} found no difference in the kinematics of the ``metal-rich'' and ``metal-poor'' stars in NGC1851.

With the only exception of NGC1851, all the clusters in our sample are very homogeneous (internally) in iron abundance, with the observed star-to-star spread fully consistent with observational errors \citep[Pap-VII]{scala}. Therefore, they seem the ideal sample for a systematic search of connections between the kinematics and the abundance of light elements known to display a significant spread within each cluster, and involved in the correlations and anti-correlations that were the main object of the overall program.
As far as we know, this exploration has never been attempted before.

All the available observational evidence on multiple populations in GCs seems to indicate that the formation of the various generations of stars and the associated chemical enrichment must have occurred at very early epochs, approximately in the first few $10^8$~yr of cluster life. 
It is therefore likely that any possible sign of a difference in the kinematical properties between different generations of stars present at those times \citep[predicted, for instance, in the models by][]{bekkimod} have been erased by two-body relaxation during the lifetime of GCs \citep[see][]{Dec08}. According to the H96 catalog, the median two-body relaxation time of Galactic GCs is just $\simeq 1$~Gyr, while the typical age is $\simeq 12.5$~Gyr. However, the same was expected for any difference in the radial distribution of various generation of stars, while it is generally observed that Na-rich/O-poor stars are more centrally concentrated than Na-poor/O-rich stars\footnote{Na-poor/O-rich stars are generally identified as belonging to the first generation of cluster stars, while Na-rich/O-poor stars are associated to the second (or any further) generation, see e.g. Pap-VII, and references therein.}\citep[see, for example,][references and discussion therein]{annib,Dec10,lardo}. Moreover, dedicated simulations are required to establish whether two-body relaxation is in fact able to remove {\em all} the signs of difference in the initial conditions over the whole radial range of the cluster \citep{lardo,bekkirot}.
Therefore, it is clearly worth verifying that any correlation between kinematic and light-element abundance is there, even if it is not expected to be observable at the present epoch. The main caveats for interpretating of our results are associated with the relatively sparse samples and by the radial distribution of the target stars that virtually misses the cluster cores (see Table~\ref{Tab_sam}, also recalling that the core radii of GCs are always smaller than their half-light radii).

   \begin{figure}
   \centering
   \includegraphics[width=\columnwidth]{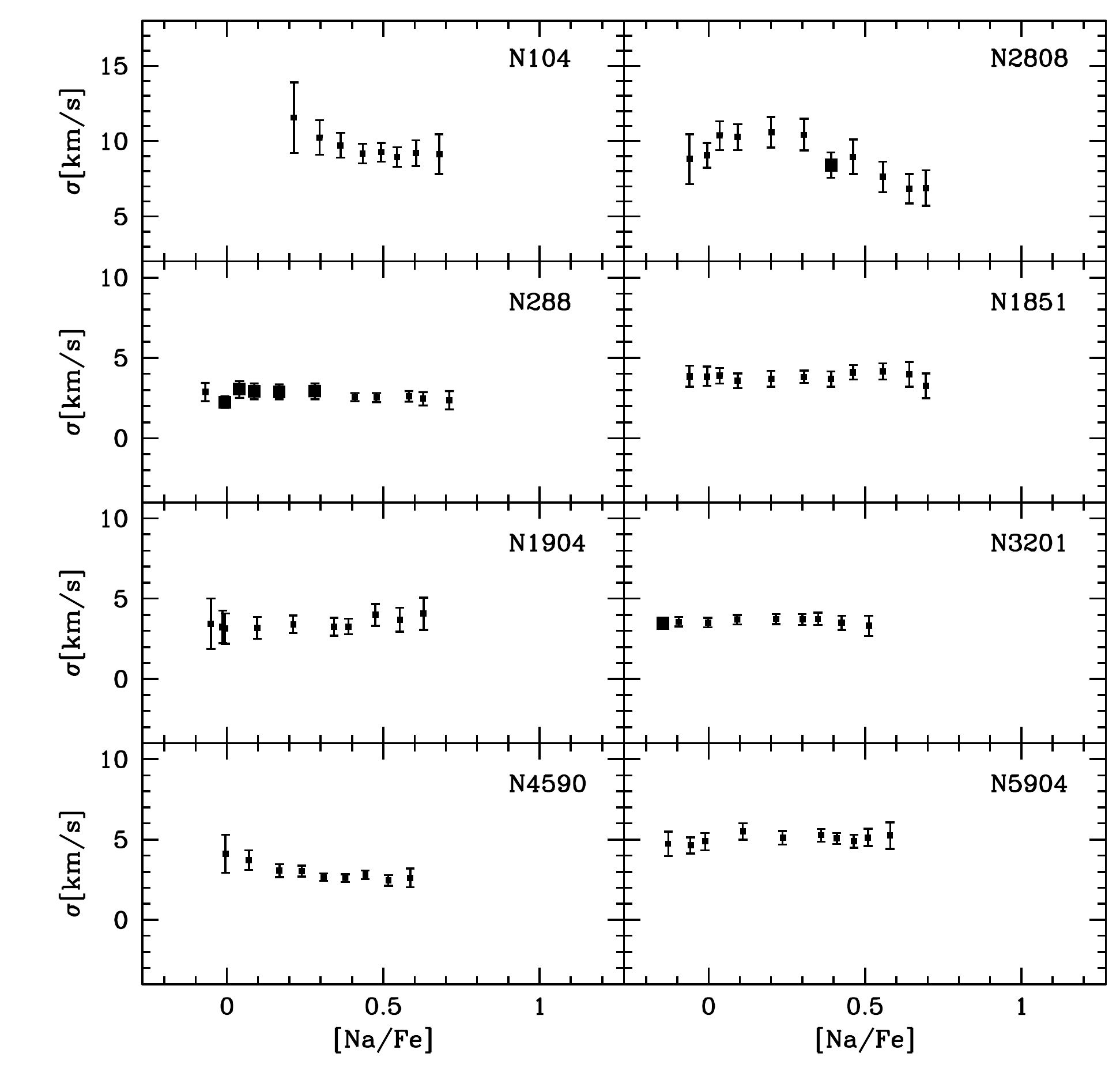}
     \caption{Velocity dispersion as a function of Na abundance for NGC104, NGC2808, NGC288,
     NGC1851, NGC1904, NGC3201, NGC4590, and NGC5904.  
     Larger filled squares indicate cases where one star has been clipped from the sample because of an RV value more than $3-\sigma$ from the mean. 
     Only estimates obtained from $\ge 10$ stars (after clipping) are plotted. 
     The vertical scale is different for NGC104 and NGC2808, to accommodate for the larger dispersion of these clusters.}
        \label{sig1}
    \end{figure}


   \begin{figure}
   \centering
   \includegraphics[width=\columnwidth]{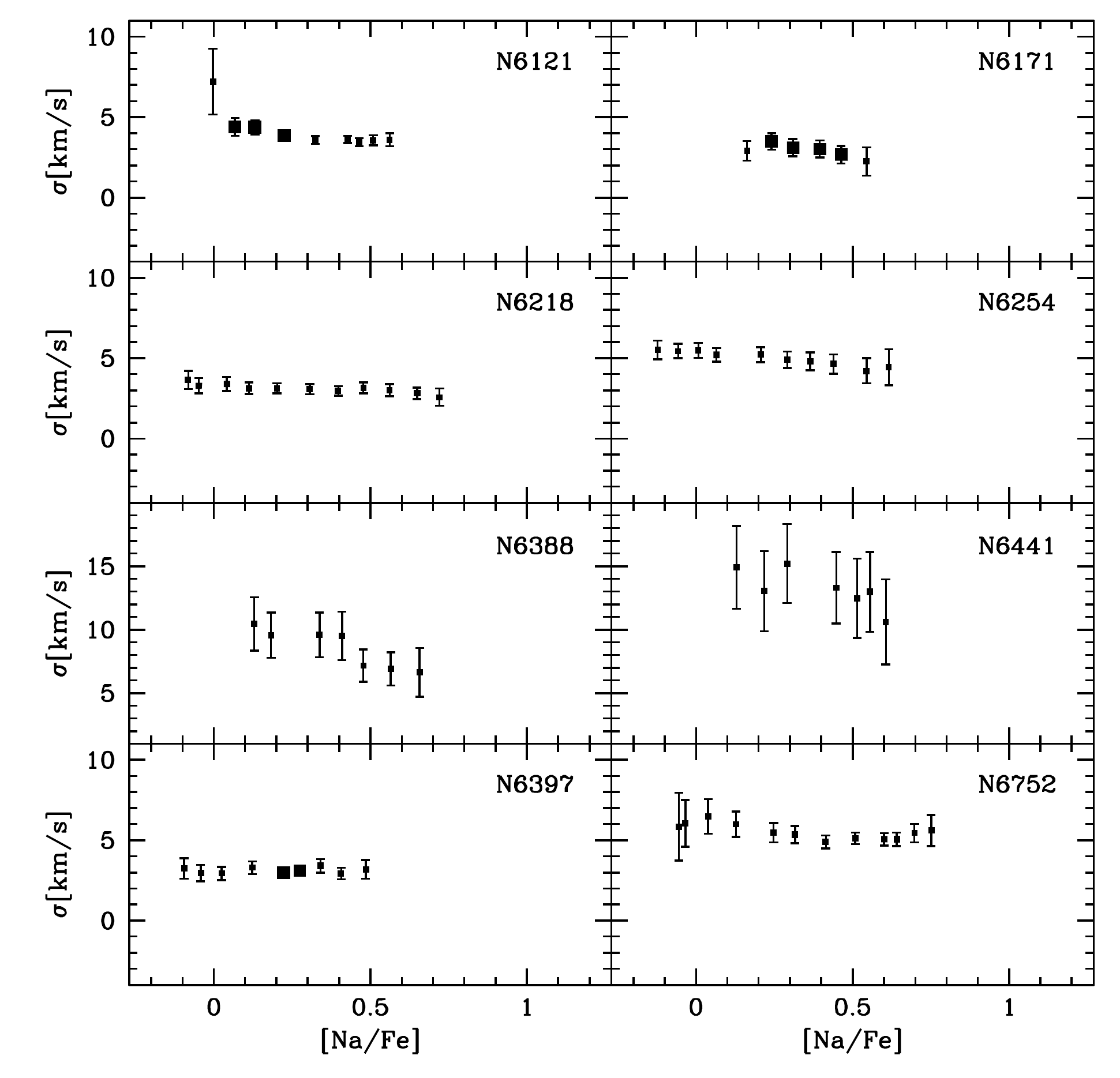}
   \includegraphics[width=\columnwidth]{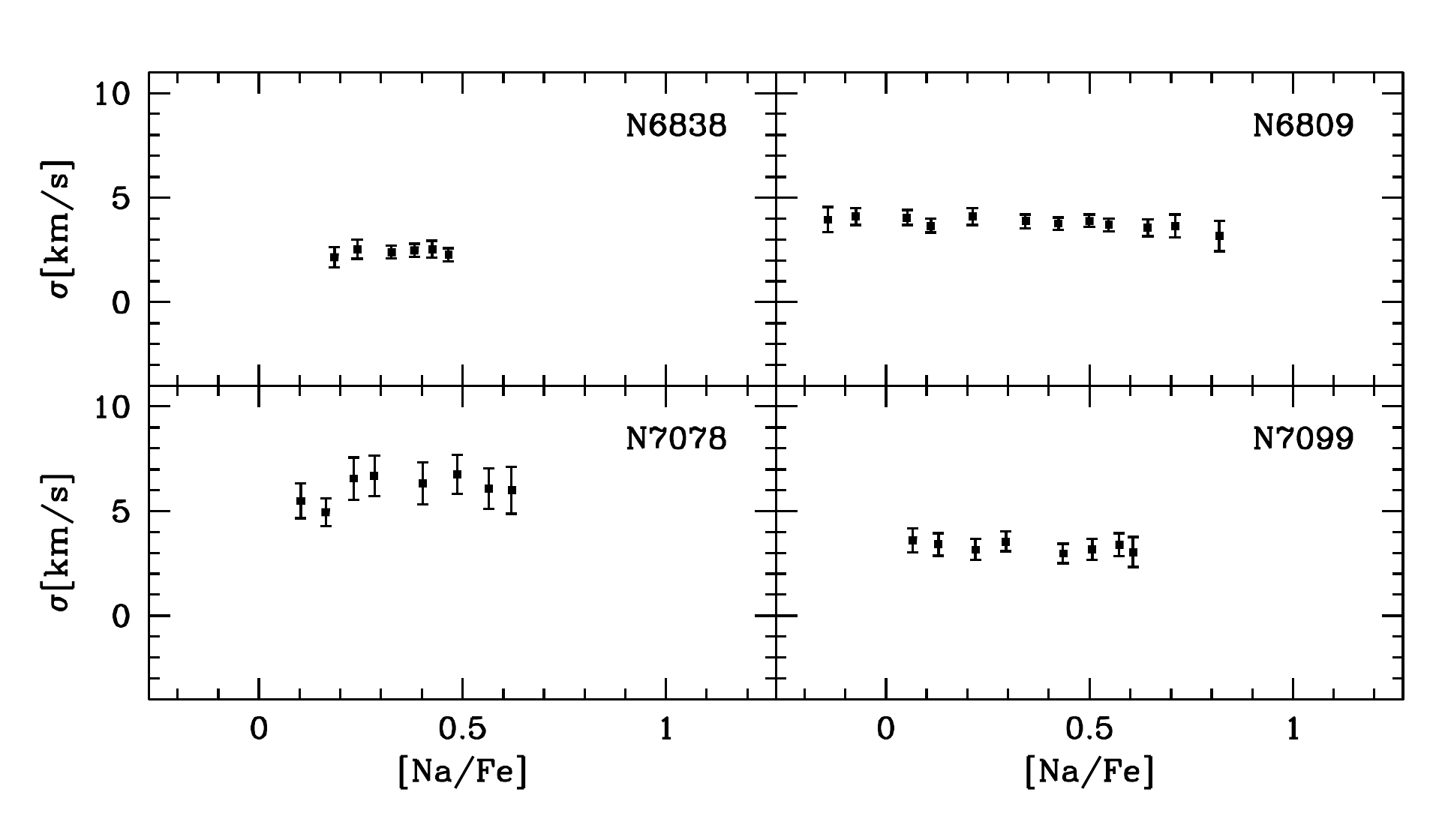}
     \caption{The same as Fig.~\ref{sig1} for the remaining clusters of our sample. The vertical scale is different for NGC6838 and NGC6441 to accommodate the larger dispersion of these clusters.}
        \label{sig2}
    \end{figure}


\begin{table*}
  \begin{center}
  \caption{Kinematics of  Na-rich$^*$ and Na-poor$^*$ stars in selected clusters}\label{Tab_F}
  \begin{tabular}{lccccccccc}
    \hline
Clus. & [Na/Fe]$_{Tr}^*$ & $\langle RV_{Na-poor}\rangle$ & $\sigma_{Na-poor}$ & $N_{Na-poor}$& 
$\langle RV_{Na-rich}\rangle$ & $\sigma_{Na-rich}$ & $N_{Na-rich}$& $F_{r/p}^a$ & Prob.$^a$\\
&&[\kms]&[\kms]&&[\kms]&[\kms]&&&\\
\hline
NGC2808 &0.4 & 103.7 &10.2& 82 & 99.4 & 7.5 &41 &1.864 & 0.008 \\
        &0.5 & 103.2 &10.0& 98 & 98.5 & 6.6 &25 &2.269 & 0.002 \\
NGC6388 &0.4 &  84.4 & 9.6& 17 & 81.1 & 7.2 &18 &1.732 & 0.132 \\
        &0.5 &  83.4 & 9.3& 24 & 81.1 & 6.6 &11 &1.957 & 0.082 \\
NGC6441 &0.4 &  19.2 &13.7& 13 & 23.6 &12.5 &16 &1.202 & 0.373 \\
        &0.5 &  22.2 &15.0& 17 & 20.8 &10.1 &12 &2.203 & 0.067 \\
NGC6752 &0.4 & -26.4 & 5.7& 51 &-26.4 & 5.0 &62 &1.302 & 0.166 \\
        &0.5 & -26.4 & 5.6& 60 &-26.2 & 5.0 &53 &1.248 & 0.202 \\
NGC6809 &0.4 & 175.0 & 4.0& 58 &174.8 & 3.6 &64 &1.263 & 0.184 \\
        &0.5 & 174.7 & 3.9& 83 &175.3 & 3.5 &39 &1.254 & 0.194 \\
\hline
\end{tabular}
\tablefoot{$^*$ The definition of Na-rich and Na-poor: Na-rich stars have [Na/Fe]$\ge$Tr, while Na-poor stars have [Na/Fe]$<$[Na/Fe]$_{Tr}$.\\
$^a$ $F_{p/r} =\frac{\sigma_{Na-rich}^2}{\sigma_{Na-poor}^2}$; Prob. = probability that the two samples 
have the same variance, according to the F-test.
} 
\end{center}
\end{table*}

As the best tracer of the  light element spread we take the sodium abundance [Na/Fe]. The main reasons for this choice are that (a) Na abundance is much easier to measure than, for instance, O abundance, and, as a consequence, reliable Na abundance estimates are available for the large majority of the stars in our sample, and, (b) Na is one of the elements showing the widest abundance spread within each globular.

In Figs.~\ref{sig1} and \ref{sig2}, we plot the velocity dispersion of cluster stars as a function of the [Na/Fe] abundance for all the program clusters. In these plots, each point corresponds to the value of the velocity dispersion $\sigma$ estimated from the stars whose [Na/Fe] is enclosed within $\pm 0.2$~dex from the central values (marked by the plotted points); these are spaced by 0.1 dex. 
 Only estimates obtained from $\ge 10$ stars (after clipping) are plotted. The errors on the velocity dispersion are computed with the jackknife bootstrapping technique \citep{lupton}, as in \citet[][B08 hereafter]{mic54}.

In the vast majority of cases dispersion curves as a function of [Na/Fe] are completely flat, within the uncertainties, so the main conclusion that can be drawn from Figs.~\ref{sig1} and \ref{sig2} is that, {\em in general, there is no correlation between the velocity dispersion and sodium abundance}.
There are only three cases that deserve some further comment. The two metal-rich and bi-modal HB morphology clusters NGC6388 and NGC6441 \citep{rich} show some marginal evidence of a drop in the velocity dispersion with increasing Na abundance, occurring around [Na/Fe]$\sim 0.4$, in line with the predictions of \citet{bekkimod} models. The same seems to  also occur for NGC2808 \citep[another cluster with multimodal HB, see][and references therein]{ema28}, but in this case the effect is more pronounced, also thanks to the larger sample and the lower uncertainties in the $\sigma$ estimates. To check the significance of the observed difference we compared the mean RV and the velocity dispersion of the Na-poor and Na-rich samples, for two different thresholds\footnote{This is {\em not} the same definition as adopted in Pap-VII to separate first and second generation stars. Here the threshold is located according to the observed change
in the velocity dispersion.} separating the two subsamples, [Na/Fe]=0.4 and [Na/Fe]=0.5, for NGC2808, NGC6388, NGC6441, and for NGC6752 and NGC6809, taken as examples of clusters with flat $\sigma$ vs. [Na/Fe] distributions. The results are reported in Table.~\ref{Tab_F}, together with the value of the F parameter ($F_{p/r} =\frac{\sigma_{Na-rich}^2}{\sigma_{Na-poor}^2}$), and the probability that the two subsamples are drawn from populations having {\em the same velocity dispersion}, according to an F test \citep{brandt}. The probability is about 20 per cent for NGC6809 and NGC6752, independent of the adopted threshold. The possibility that the two subpopulations have the same $\sigma$ is clearly far from excluded, for NGC6388 and NGC6441, but the probability is lower than ten per cent in both cases, at least for the [Na/Fe]=0.5 threshold. Hence, the hint of a difference shown here may need to be explored with larger samples. On the other hand, the probability is lower than one per cent in the case of NGC2808, independent of the adopted threshold. The NGC2808 sample is rich enough to allow for a more detailed investigation.

\subsubsection{The case of NGC2808}
\label{case}

\begin{figure}
   \centering
   \includegraphics[width=\columnwidth]{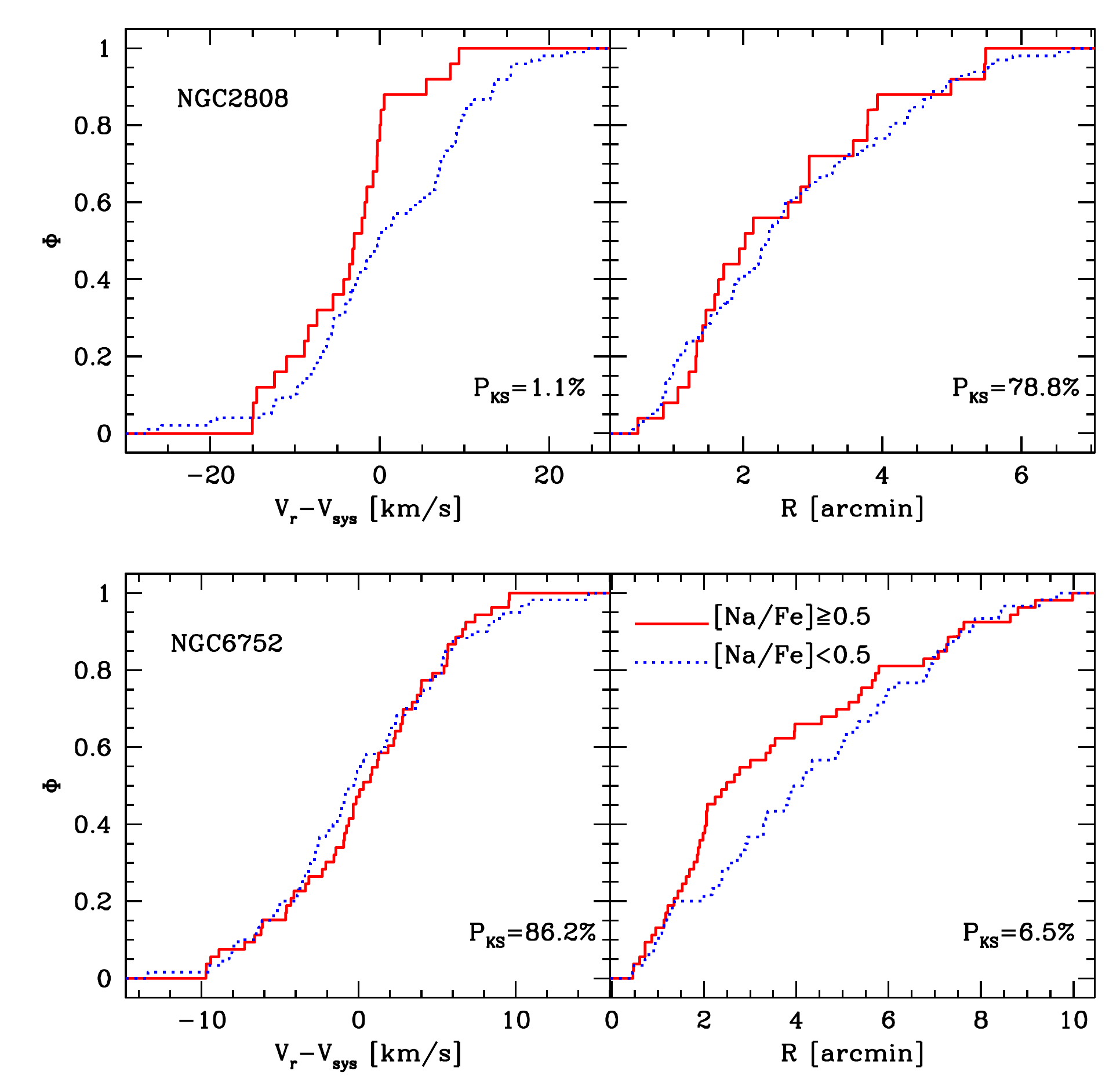}
     \caption{Comparison of cumulative distributions for Na-rich (red continuous line) and Na-poor (blue dotted line) for NGC2808 (upper panels) and NGC6752 (lower panels). Left panels: distributions of the radial velocity in the cluster system. Right panels: distributions of the distances from the cluster center. The probability that the Na-rich and Na-poor populations are drawn from the same parent population, according to a Kolmogorov-Smirnov test, is reported in the lower right corner of each panel. The adopted threshold between Na-poor and Na-rich stars is [Na/Fe]=0.5.} 
        \label{track}
    \end{figure}

An obvious possibility for the origin of the lower $\sigma$ of Na-rich (w.r.t. Na-poor) stars in NGC2808 is that the former may have a radial distribution less concentrated than the latter. In any isotropic mass-follow-light self-gravitating equilibrium system, the velocity dispersion reaches its maximum value at the center and declines with radius. Independently of anisotropy, this is also observed to be the case for NGC2808 (see below). Although Na-rich stars are usually observed to be {\em more} centrally concentrated than Na-poor stars, still an unlucky selection of targets may turn out in a Na-rich sample typically lying at larger radii, thus implying a lower $\sigma$ with respect to Na-poor. 
To check this possibility we plot in Fig.~\ref{track} the RV and radial cumulative distributions for Na-rich and Na-poor stars in NGC2808 and NGC6752, for comparison.

In NGC6752 the two RV distribution are indistinguishable, while the higher concentration of the Na-rich sample is obvious, and likely tracing a real difference in the radial distribution of the two populations \citep[see][]{IQR,lardo}. In NGC2808 we see the opposite case: the radial distributions are indistinguishable, while the RV distribution of Na-rich stars is much steeper than for Na-poor. According to a Kolmogorv-Smirnov (KS) test, the probability that the two subsamples are from the same parent distribution of RV is just 1.1 per cent, in excellent agreement with the result of the F test, so the observed difference in $\sigma$ does not come from a difference in radial distribution 
(the probability that the velocity distribution of Na-rich and Na-poor stars of NGC2808 are drawn from the same parent population decreases to 0.2\% if  [Na/Fe]=0.4 is adopted as threshold). 
However the upperleft hand panel of Fig.~\ref{track} provides an interesting hint in this sense. It can be readily appreciated that nearly 90 percent of the Na-rich subsample lies at negative velocity, in the reference system of the cluster (in the following RV or $V_r$ have always to be intended as expressed in the reference system of the clusters, i.e. with the systemic velocity of the clusters subtracted). This is confirmed by looking at the mean velocities of the two samples, which are observed to differ by $\sim 4.0$~\kms.  
 
\begin{figure*}
   \centering
   \includegraphics[width=\columnwidth]{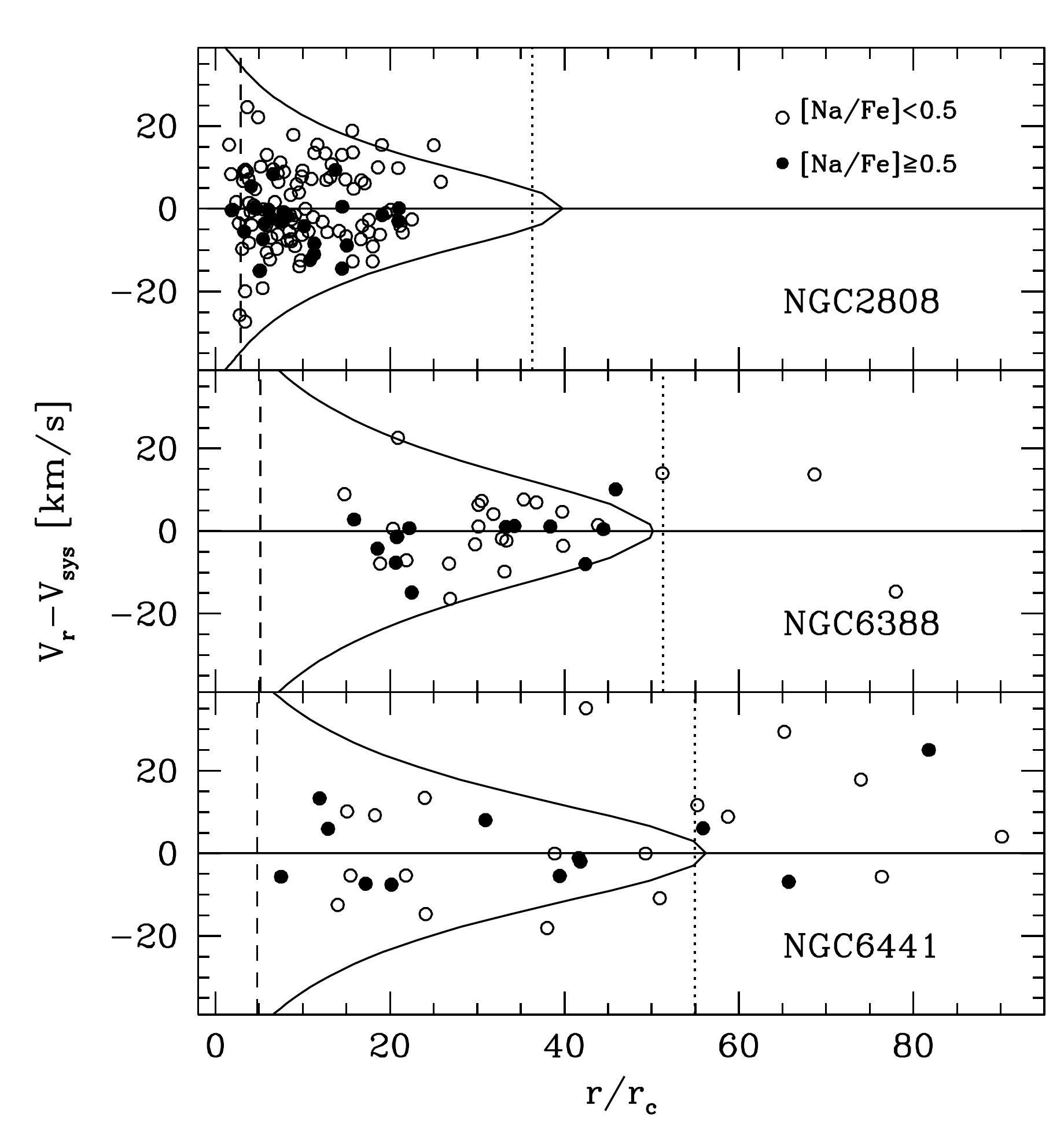}
   \includegraphics[width=\columnwidth]{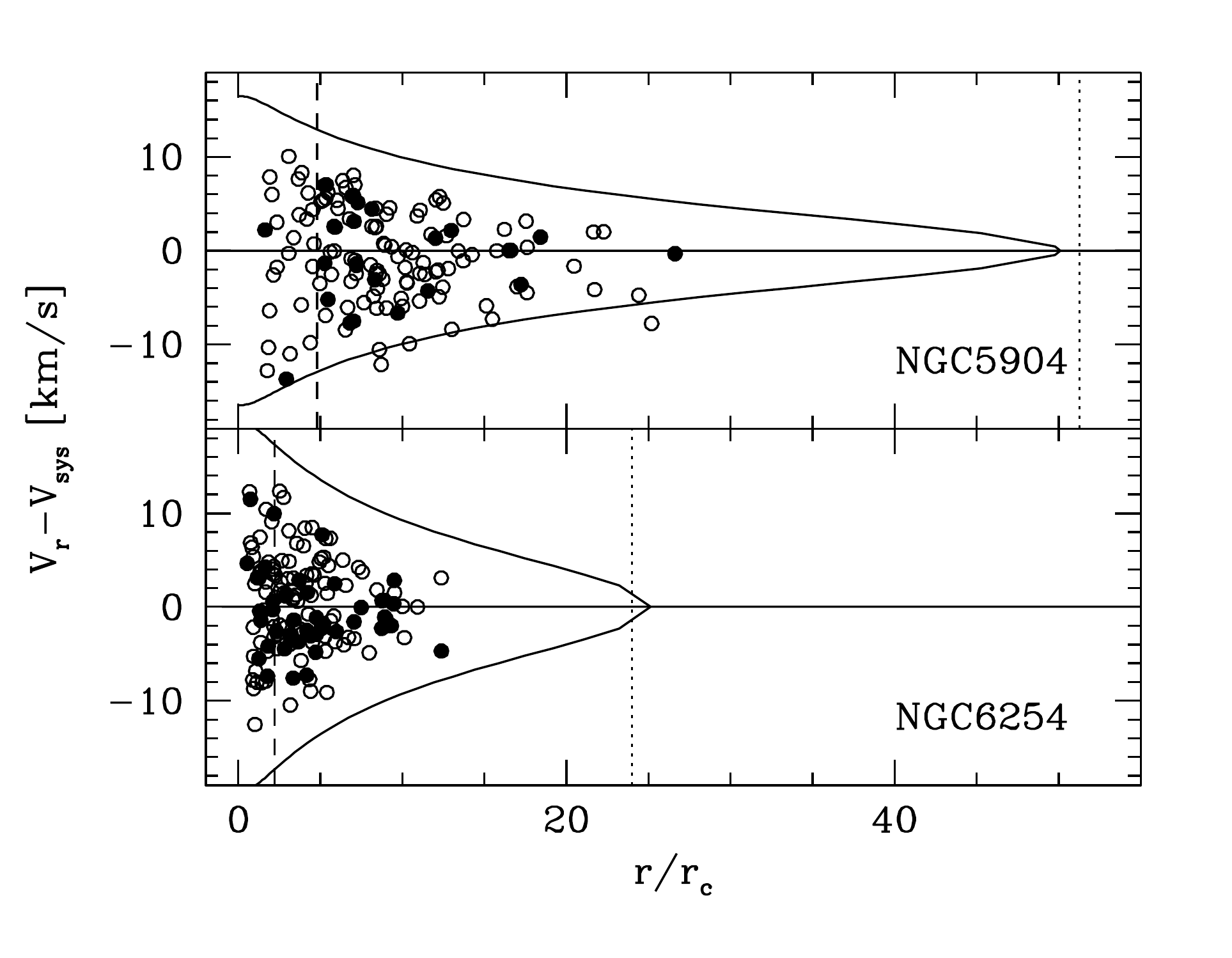}
     \caption{Radial velocity in the cluster system as a function of distance from the cluster center, expressed in core radii units, for less studied clusters. Na-rich and Na-poor stars are plotted with different symbols. The dotted line marks the tidal radius of the K66 model that best fits the surface brightness profile, according to \citet{mcv}. The continuous curves are the $\pm 3\sigma$ profiles of the same models, normalized to the central $\sigma$ listed by \citet{h96}, approximately representing the envelope of the allowed velocities for the bound members of the clusters. The dashed line marks the half-light radius, taken from \citet{h96}.}
        \label{cosig1}
    \end{figure*}

In the uppermost panel of Fig.~\ref{cosig1}, it is apparent that the distribution of RV with radius separates into two branches around $r\sim 5~r_c$, as if the velocity distribution was bimodal at large radii. This is strongly suggestive of systemic rotation, as is indeed confirmed in Sect.~\ref{rot}. Moreover, it is clear that the large majority of the Na-rich stars belongs to only one of the two branches. In Sect.~\ref{rot} we will show that the two branches corresponds to the two wings of the cluster rotation curve. Likely by mere chance, our sample contains more stars from the half of the cluster that is approaching than from the half that is receding. This asymmetry in the distribution of targets has been exacerbated in the Na-rich sample, enough to produce the significant difference detected by the F and KS tests (that assume no selection effects, i.e. even sampling) but still compatible with a chance occurrence.

In our view, there are two interesting conclusions that can be drawn from this case. First, it would be worth verifying this result with a larger and more evenly distributed sample. While it is likely that the observed effect is from the described rotation+sampling conspiracy, we cannot exclude that there is a real difference. If this were the case, it would have a striking impact on any scenario for the formation of this very peculiar cluster \citep{2808,p28,angie28,amaro}. Second, it provides an example of the possible dangers connected with neglecting rotation in the analysis of velocity of stars in a GC, when the amplitude of the rotation ($A_{rot}$, see Sect.~\ref{rot}) is relatively high with respect to the dispersion ($A_{rot}/\sigma_0\ga 0.2$).

\subsection{Rotation and Na abundance}
\label{rona}

A distinctive prediction of  the models for the formation of GCs with multiple populations by 
\citet{bekkirot,bekkimod} is that even if a very small fraction ($s$=0.3 percent) of the kinetic energy of the gas forming the first generation (FG) of stars is in form of bulk rotation energy, stars in the second generation (SG) acquire significant rotation, much stronger than what remains in the FG. In particular, $s$ values in the range 0.3 -18.0 percent correspond to ratios of rotational velocity to velocity dispersion ($V_{rot}/\sigma$) from 0.75 to 2.5 for FG stars. \citet{bekkirot} concludes that {\em ``... the rotation of SG
{\rm [stars, ndr]} is due largely to dissipative accretion processes of AGB ejecta from FG with initially a small amount of rotation ...''}. 
In Bekki's model the formation of both generations of stars is essentially completed after a few $ 10^8$~yr from the formation of the very first star, so, this possible difference in the rotation of different generations of cluster stars should  also have been largely smoothed out by subsequent dynamical evolution, during the $>10$~Gyr of a cluster life \citep[see][]{Dec08}. Nevertheless, dedicated simulations are required to check in detail whether some fossil of such difference may still be observable in GCs at the present epoch, given the initial conditions considered in \citet{bekkirot}. On the other hand, significant rotation of the cluster as a whole is a general outcome of the merging of two clusters \citep{makino}, another channel that has been suggested as playing a role in the formation of multiple populations GCs 
\citep[][and references therein]{n1851,amaro}.

We searched for differences in the rotation amplitude between Na-rich ([Na/Fe]$\ge 0.4$) and Na-poor 
([Na/Fe]$< 0.4$) stars within each program cluster using the same kind of analysis adopted and described in Sect.~\ref{rot}, below, to look for rotation in the total samples. We cannot reach definitive conclusions from our data, since this kind of analysis would require larger samples to allow a fully reliable study of {\em subsamples}, and we lack coverage of the innermost regions of the clusters, where most of the SG population is expected to reside in Bekki's model. In addition, rotation amplitudes are intrinsically small in GCs (L09, L10a,b), and we were forced to consider only the [Na/Fe]=0.4 threshold to avoid too poorly populated Na-rich subsamples. It is also worth noting that Na features are among the most prominent ones for metal-poor stars in the spectral range covered by our data. Therefore stars with low Na abundance (i.e. weak Na features) may have slightly larger uncertainties in RV than stars with high Na abundance.

We simply report that {\em we do not find any significant difference (i.e. $\ga 2$~\kms) in mean rotation amplitude ($A_{rot}$, see below) between Na-rich and Na-poor stars} for cluster having at least 20 stars per subsample\footnote{The observed differences in the position angle of the rotation axis maximizing the rotation amplitude ($PA_0$, see Sect.~\ref{rot}) are not considered as reliable, because of the large uncertainties associated with the estimate of this parameter from such small samples.}. Furthermore, any marginal difference is in the opposite sense with respect to Bekki's model expectations; i.e., in general, Na-poor samples tend to have larger rotation amplitudes than Na-rich samples. 

In NGC6171 the Na-poor group displays an amplitude of 5.5~\kms, but the Na-rich one only 0.5~\kms. The two subsamples contain only 20 and 13 stars, thus we should not trust this result at face value.
In NGC7078 the Na-poor group displays an amplitude of 6.5~\kms, and the  Na-rich one just 1.0~\kms, but the two subsamples contains only 20 and 31 stars; in addition, this is the cluster having the worst quality RV estimates in our survey. It is also worth noting that Na features are among the most prominent ones for very metal-poor stars in the spectral range covered by our data. Therefore stars with low Na abundance (i.e. weak Na features) may have larger uncertainties in RV than stars with high Na abundance, which may be relevant in the case of NGC7078 where the overall RV uncertainties are quite large.

Finally a brief comment is needed on the clusters showing signs of Na-abundance vs. $\sigma$ correlation. 
The samples for NGC6441 and NGC6388 are as small as the one for NGC6171; however, the Na-rich and Na-poor groups display very similar rotational patterns in both clusters (see below).
In NGC2808, where the Na-rich and Na-poor groups contain 41 and 82 stars, respectively, Na-poor stars display a larger rotational amplitude than Na-rich stars ($A_{rot}\simeq 4.5$~\kms and $A_{rot}\simeq 2.0$~\kms, respectively). This is likely accounted for by the fact that the majority of Na-rich stars lie mainly on one of the rotation curve wings, as anticipated above and discussed in Sect.~\ref{rot}.

\section{Dispersion and rotation for the less studied clusters}
\label{disp}

There are a few clusters among those listed in Table~\ref{Tab_sam} for which the present sample can provide a significant improvement over existing kinematic data and analysis. For NGC2808, NGC6388, and NGC6441, only estimates of $\sigma_0$ from integrated spectra are available. For NGC5904, in addition to the integrated measure, a sample of RV for 46 individual stars was considered by \citet{rs91}, still significantly smaller than our sample (136 stars). Finally, NGC6254 lacks any integrated measure, the only existing estimate of the velocity dispersion for this cluster coming from RVs for 25 stars from \citet{rs91} in the range $0.6r_c - 6.9r_c$, while our sample includes 147 stars. In the following we briefly discuss the properties of the RV distributions in these clusters, providing new estimates of 
$\sigma_0$ if we consider the case worth it (see below). 

In Fig.~\ref{cosig1} we show the RV distribution in the cluster system as a function of projected distance from the cluster center, expressed in units of core radius. We superposed the $\pm 3.0\sigma$ contour of the K66 model that is found by MvM05 to best fit the surface brightness profile of the clusters, as a reference. The position of $r_h$ and $r_t$ are also reported\footnote{The outer edge of the K66 profiles does not exactly match the reported $r_t$ (dotted line). This is because we selected the K66 model having the nearest $C=log(r_t/r_c)$ to the value reported by \citet{mcv} from a grid of models sampling the range of C at 0.1-0.05 steps. For example, for NGC2808, \citet{mcv} report C=1.56 and we adopted the model having C=1.6.}. Na-rich and Na-poor stars are plotted with different symbols.

At first glance, Fig.~\ref{cosig1} reveals both (a) the (already noted) poor sampling of the innermost regions of the clusters provided by our data, and (b) the wide variety in the radial sampling from cluster to cluster. For example, the NGC6254 sample has a significant fraction of stars within $r_h$ and barely reaches $0.5r_t$, while the innermost stars of the NGC6388 and NGC6441 lie around $\ga 2r_h$, and the farthest are far beyond the tidal radius. As discussed in the introduction our samples are not best suited to a full characterization of the clusters kinematics. Still, in several cases, they are competitive with the best samples available in the literature, and, for the five clusters considered in this section, they are clearly superior to other available samples.

The adopted K66 models appear to provide a reasonable description of the observed velocity distribution for  NGC6254, NGC2808, and NGC5904. In the last two cases there are some stars lying beyond the $\pm 3.0\sigma$ contour of the best-fit model. This may be due to several trivial factors: uncertainty in the adopted value of $\sigma_0$ providing the normalization constant of the model profile, uncertainties in the parameters of the best-fit model and/or limitations in the available data for the surface brightness profile \citep[see, for example][and references therein]{correnti}, and inadequacy of the K66 model in representing the kinematics of the clusters (e.g. rotation, anisotropy, see MvM05). In addition to these issues, recent detailed studies have pointed out that a large fraction of unbound (due to Galactic tides) cluster stars can contaminate the velocity dispersion profiles of actual GCs well inside their Jacobi radius. For example, in the $M=10^4~M_{\sun}$ models considered by \citet{kupper}, unbound stars are found to dominate the samples beyond a radius $\simeq 0.5$ of the Jacobi radius.  

The NGC6388 and NGC6441 samples are too sparse to let us draw firm conclusions. At face value they show a  pretty flat dispersion curve over the considered radial range. This, along with the significant rotation (see below) and the presence of several members beyond the K66 tidal radius, may be suggestive of strong effects from Galactic tides \citep[][and references therein; see also the discussion in Sect.~\ref{summ}, below]{munoz,sollima,kupper}. The large number of members beyond the tidal radius seen in NGC6441 may simply be due to the inadequacy of K66 to describe the outer regions of clusters that is discussed in detail by MvM05. These authors found that in most cases spherical and isotropic W75 models (more extended than K66 ones) provide the best overall fit of cluster observed light profiles. This is actually the case for NGC6441, while for NGC6388, K66 is better.
The tidal radii for the best-fit W75 models are 4.2 times and 6.1 times larger than their K66 counterparts, for NGC6388 and NGC6441, respectively. All the stars in our samples would thus lie well within the W75 tidal radius of the clusters.

\subsection{New $\sigma_0$ estimate for NGC5904}

   \begin{figure}
   \centering
   \includegraphics[width=\columnwidth]{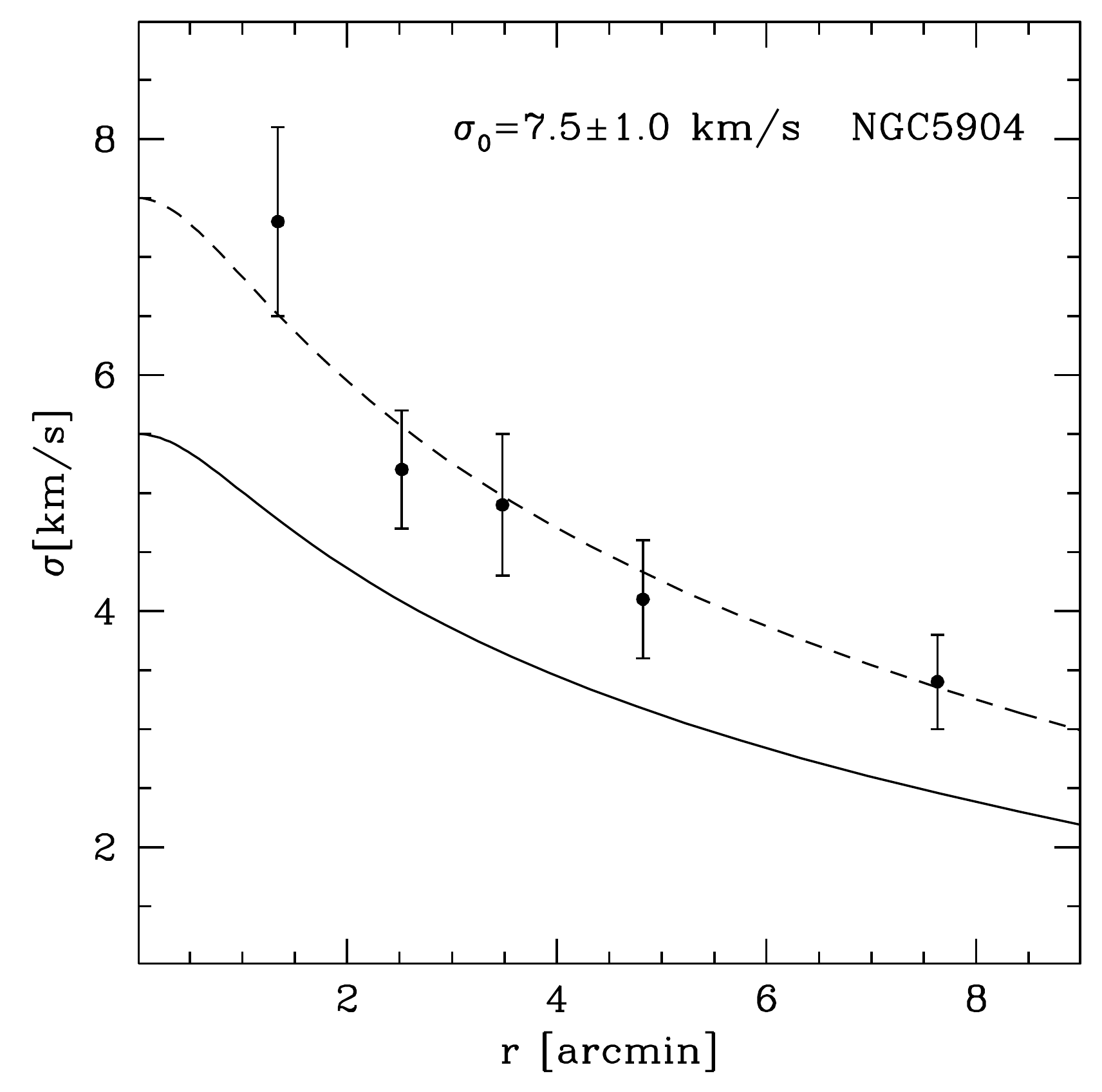}
     \caption{Comparison of the observed velocity dispersion profile for NGC5904 as derived from our data (filled circles with error bars) and the prediction of the K66 model that best fits the SB profile, according to MvM05, with $\sigma_0$ taken from H96 (thick continuous line). An alternative value of $\sigma_0$, providing significantly better representation of the observed profile, is proposed ($\sigma_0=7.5$~\kms, from our own fit, dashed line). }
        \label{sip}
    \end{figure}


In the comparisons with other self-gravitating systems, the kinematic properties of a GC are usually summarized by a single number, i.e. the central velocity dispersion $\sigma_0$ \citep{djorg,mfp}. This is the number usually reported in GC catalogs (PM93, H96), so it may be useful to provide refinements of such estimates or to confirm existing estimates on significantly sounder basis, when possible. 

We used our data to produce velocity dispersion curves for the less studied clusters listed above, in the same way as described in B08, using jackknife resampling to compute uncertainties. 
We compared the observed curves with the predictions of the K66 models plotted in Fig.~\ref{cosig1}, normalized with the $\sigma_0$ listed in H96, to check that our new results are compatible with these values. 

The agreement is acceptable or satisfying for NGC6388, NGC2808, and NGC6254, so we confirm the $\sigma_0$ estimate reported in the literature for these clusters. 
From the sparse sample of NGC6441 we were able to obtain only one point of the dispersion curve within the tidal radius of the best-fitting K66 model; this point is compatible with the listed value of $\sigma_0$. Outside of $r_t$ the velocity dispersion remains fairly large ($\sigma\sim 10$~\kms).

A direct comparison of the observed velocity dispersion profiles with the predictions of the K66 models is plotted in Fig.~\ref{sip} for NGC5904, the only case where we found a mismatch between the current normalization of the velocity profile of the model and our data. The value listed by H96 ($\sigma_0=5.5\pm 0.4$~\kms, see Table~\ref{Tab_sam} and PM93 for the original data sources) is clearly too low to match the observed dispersion curve. The adopted $\sigma_0=7.5\pm 1.0$~\kms~ instead provides a satisfying fit, and we propose this value as a new, more robust estimate of the central velocity dispersion for this cluster. It has to be recalled that the accuracy of this estimate also depends on how good and appropriate the adopted K66 models are (see MvM05).  

   \begin{figure}
   \centering
   \includegraphics[width=\columnwidth]{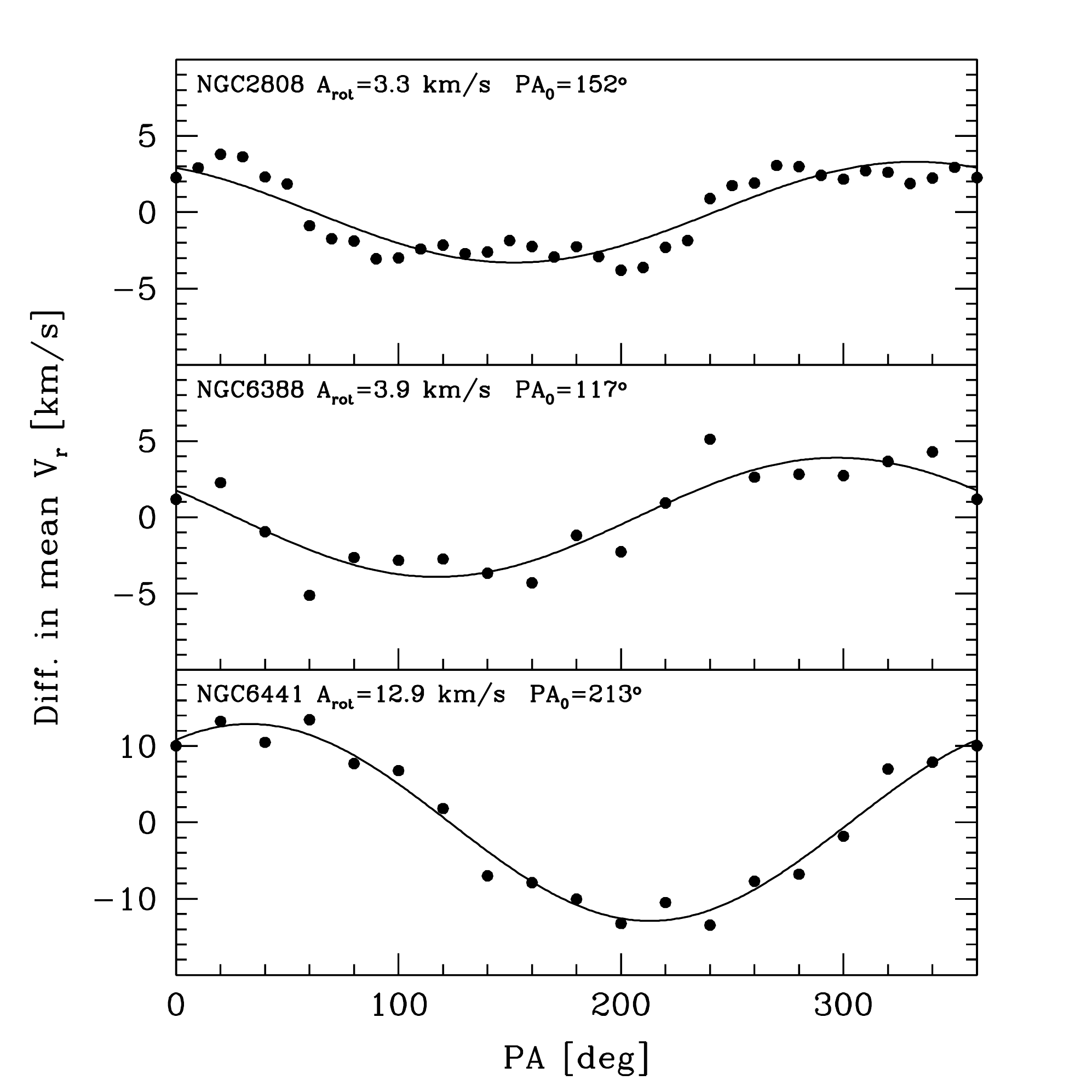}
     \caption{Rotation in NGC2808, NGC6388, and NGC6441.
     The plots display the difference between the mean velocities on each side of a cluster with respect to    a line passing through the cluster center with a position angle PA (measured from north to east, north=0$\degr$, east=90$\degr$), as a function of the adopted PA. The adopted line coincides with the projected rotation axis. The continuous line is the sine law that best fits the observed pattern. The vertical scale used for NGC6441 is larger, to accommodate the strong rotation signal. It is also worth noting that even limiting the NGC6441 sample to the 21 stars lying within the tidal radius (see Fig.~\ref{cosig1}) the rotation pattern is very similar, with $A_{rot}\simeq11$~\kms and  PA$_0\simeq$225$\degr$.}
        \label{corot1}
    \end{figure}


\subsection{Rotation}
\label{rot}

Globular clusters are generally considered as classical examples of pressure-supported systems, in which rotation is nonexisting or negligible. The classical case of rotating GC is $\omega$~Cen,  which has been considered as very peculiar in many aspects for a long time, and it displays a 
large-amplitude ($\sim 8$~\kms) rotation pattern coupled with high isophotal ellipticity \citep{mmm97,vandeven,sollima}. However, recent studies based on large samples are revealing that rotation of typical amplitude $2-4$~\kms ~is far from uncommon in Galactic GCs \citep[see][L09, L10a,b]{cote,ak03,bosch}. Rotation amplitudes detected from RV samples are just lower limits to the true 3D amplitude, because of projection onto the plane of the sky. This may support the suggestion by MvM05 that W75 models may be more appropriate for describing actual GCs, as these models can account for rotation \citep[see][for a recent application; however, Wilson's models adopted by MvM05 do not include rotation]{sollima}.

We used our sample to search for rotation in all the clusters for which this characteristic has not been considered in previous studies\footnote{For the remaining clusters we note that the results of our rotation analysis are in good agreement with those reported in the literature and based on larger samples, in most cases from L10b.}. In addition to these, we also analyzed NGC1851 and NGC1904, whose rotation has been considered by \citet{sca11}, but only separated in two radial bins, while we aim at global properties to make proper comparisons among clusters. Our results are in good agreement with those by \citet{sca11}. To get a rotation amplitude value that is homogeneous with the other clusters ,we also studied the case of NGC7078, whose kinematics have been analyzed in much greater detail by \citet{bosch}.
We limit our analysis to rotation patterns that can be detected from radial velocities, since the data on rotation in the plane of the sky are only available for a handful of clusters 
\citep{ak03,vandeven,bosch}. 

   \begin{figure}
   \centering
   \includegraphics[width=\columnwidth]{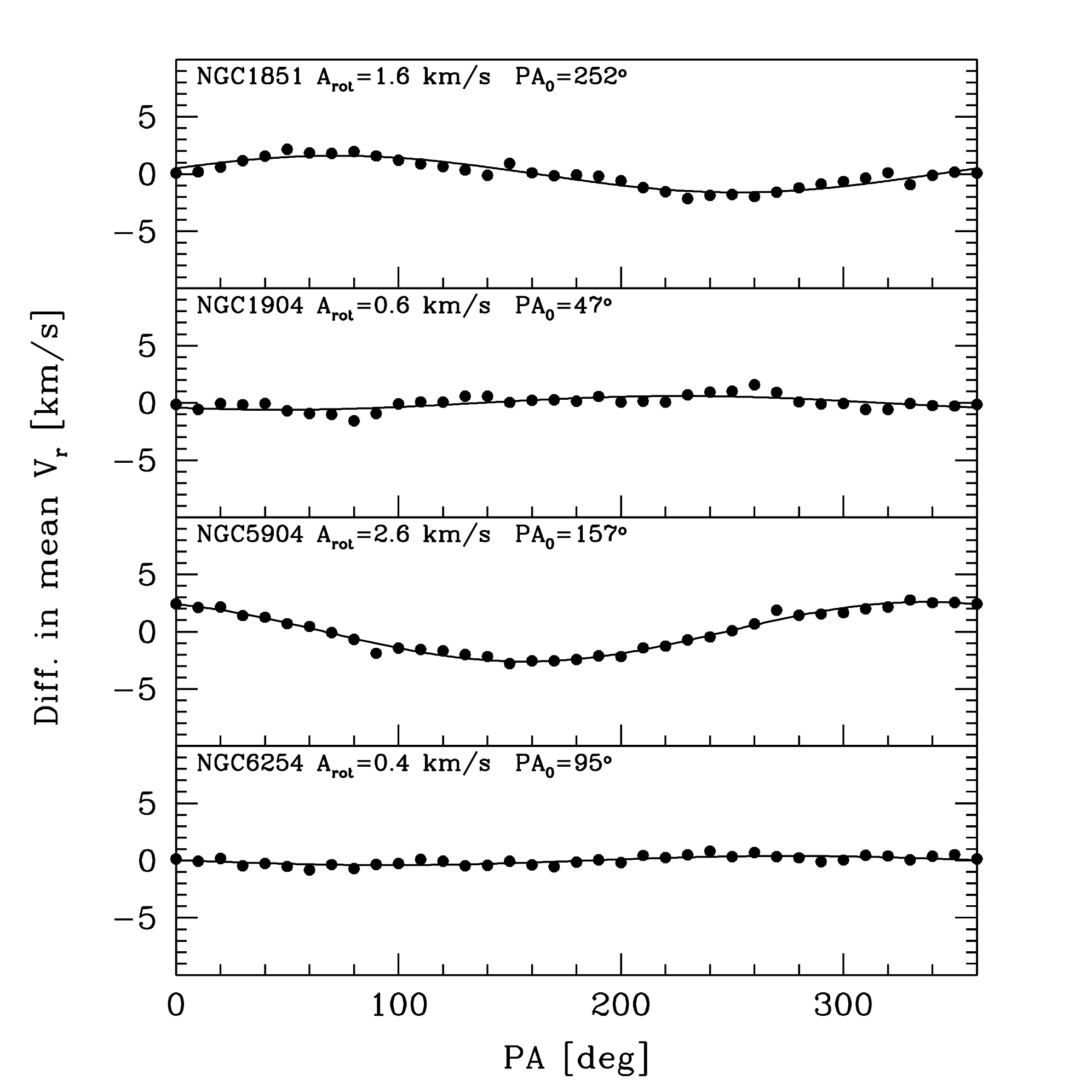}
   \includegraphics[width=\columnwidth]{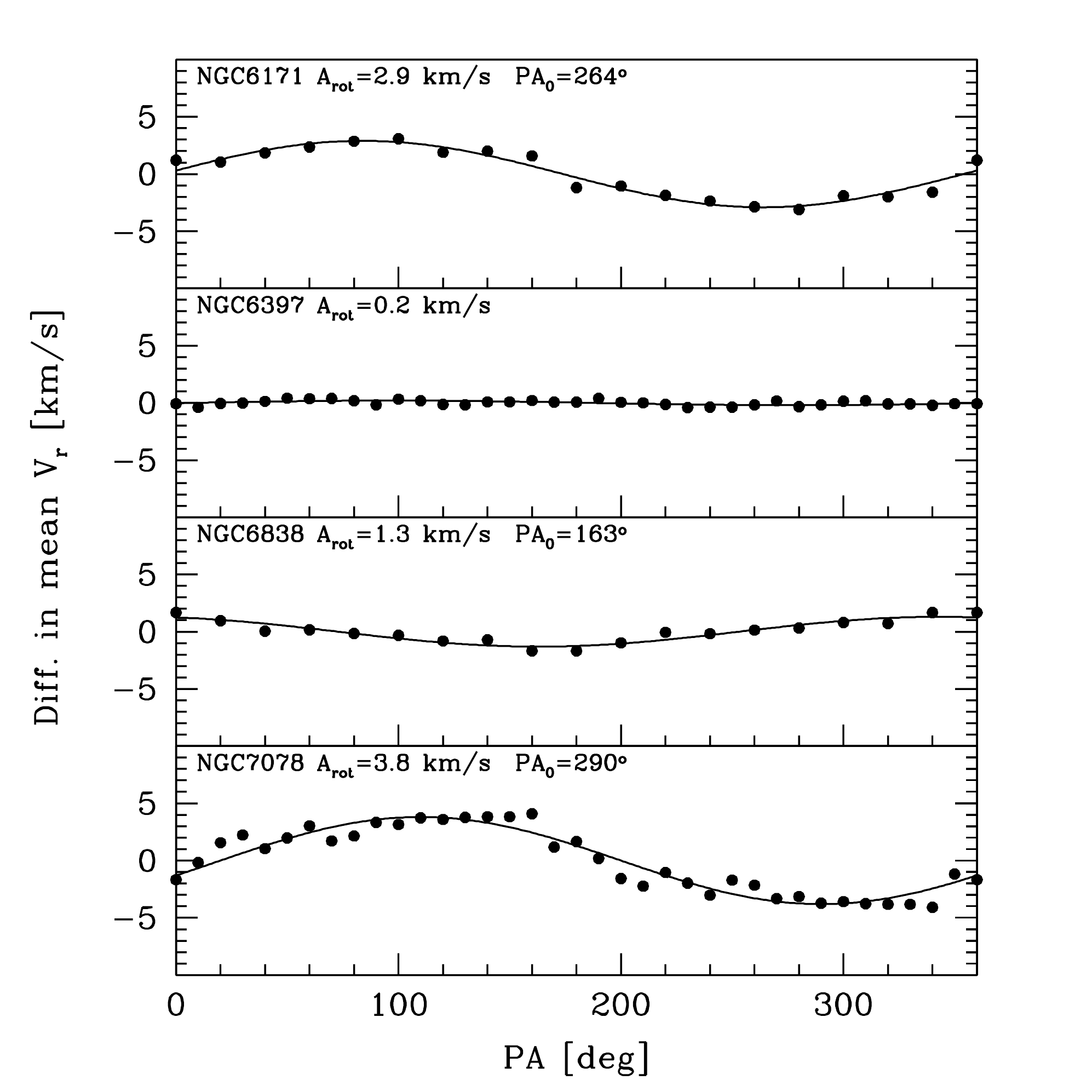}
     \caption{The same as Fig.~\ref{corot1} but for NGC1851, NGC1904, NGC5904, NGC6254, NGC6171, NGC6397, NGC6838, and NGC7078. The kinematics of some of these clusters has been studied
     with samples larger than ours, still they lack a study of rotation (NGC6171, NGC6397, NGC6838) or an estimate of the rotation amplitude homogeneous with those obtained here.}
        \label{corot2}
    \end{figure}




To look for rotation we used the same method as was adopted by \citet{cote}, \citet{panci07}, L09, and L10a,b.
For any given cluster, we divided the sample in two, choosing stars lying on the opposite side of a line passing from the cluster center, and we computed the difference in mean RV between the two subsamples. Then we rotate the dividing line by a fixed amount ($10\degr$ or $20\degr$, depending on the number of stars of the considered sample) and repeated the computation. Recording the difference in the mean RV for each position angle (PA) of the dividing line we can plot the first as a function of the second: a coherent sinusoidal pattern is a clear sign of rotation. In the adopted approach PA is defined to increase anti-clockwise in the plane of the sky from north (PA=0$\degr$) toward east (PA=90$\degr$).
We fit the observed patterns with the simple sine law:

\begin{equation}
\Delta \langle V_r\rangle = A_{rot}sin(PA+\Phi)
\end{equation}

\noindent
 where $\Phi=270\degr-PA_0$, and $PA_0$ is the position angle of the dividing line corresponding to the maximum rotation amplitude (in degrees), coinciding with the rotation axis, and $A_{rot}$ is two times the actual mean amplitude (in ~\kms; see L10a). However, it must be noted that, as the $\langle RV\rangle$ difference is averaged over the full range of radius covered by the sample, $A_{rot}/2$ should be an underestimate of the maximum rotation amplitude, which should be the most informative global rotation parameter \citep{MH97}, since the amplitude generally varies significantly with distance from the cluster center \citep[see, for example, the case of $\omega$~Cen][]{sollima}. By inspecting some well-populated rotation curves of GCs, we noted that $A_{rot}$ is a reasonabl proxy for the actual maximum amplitude \citep[see][]{panci07}; in the following we use this value as an estimate of the rotation amplitudes and, consequently multiply by two the values reported by L10b, to make the two sets of measures fully homogeneous. We explored the reliability of the derived parameters by repeating the analysis on subsamples and by comparing it with external samples (e.g. for clusters in common with L11), and we concluded that the estimates of $A_{rot}$ are fairly robust (we assume a typical $1\sigma$ uncertainty of 0.5~\kms  ~for the largest samples, similar to L10b), while $PA_0$ is quite sensitive to the adopted sample and should be considered as uncertain at the $\pm 30\degr$ level in the best cases.

The results of the analysis are displayed in Figs.~\ref{corot1} and \ref{corot2}. It is evident  that the considered clusters cover a wide range of rotation properties, from no rotation (NGC6397) to an amplitude of more than 10~\kms ~(NGC6441). It is interesting to note that three among the most peculiar clusters in terms of multiple populations, NGC2808, NGC6388, and NGC6441, display particularly strong rotation patterns, possibly hinting at a similarity with some galactic nuclei \citep{seth}. Other clusters for which we detect for the first time a significant amplitude of rotation are NGC5904 and NGC6171.
The significant amplitude found for NGC7078 agrees with the results by \citet{bosch}.

   \begin{figure}
   \centering
   \includegraphics[width=\columnwidth]{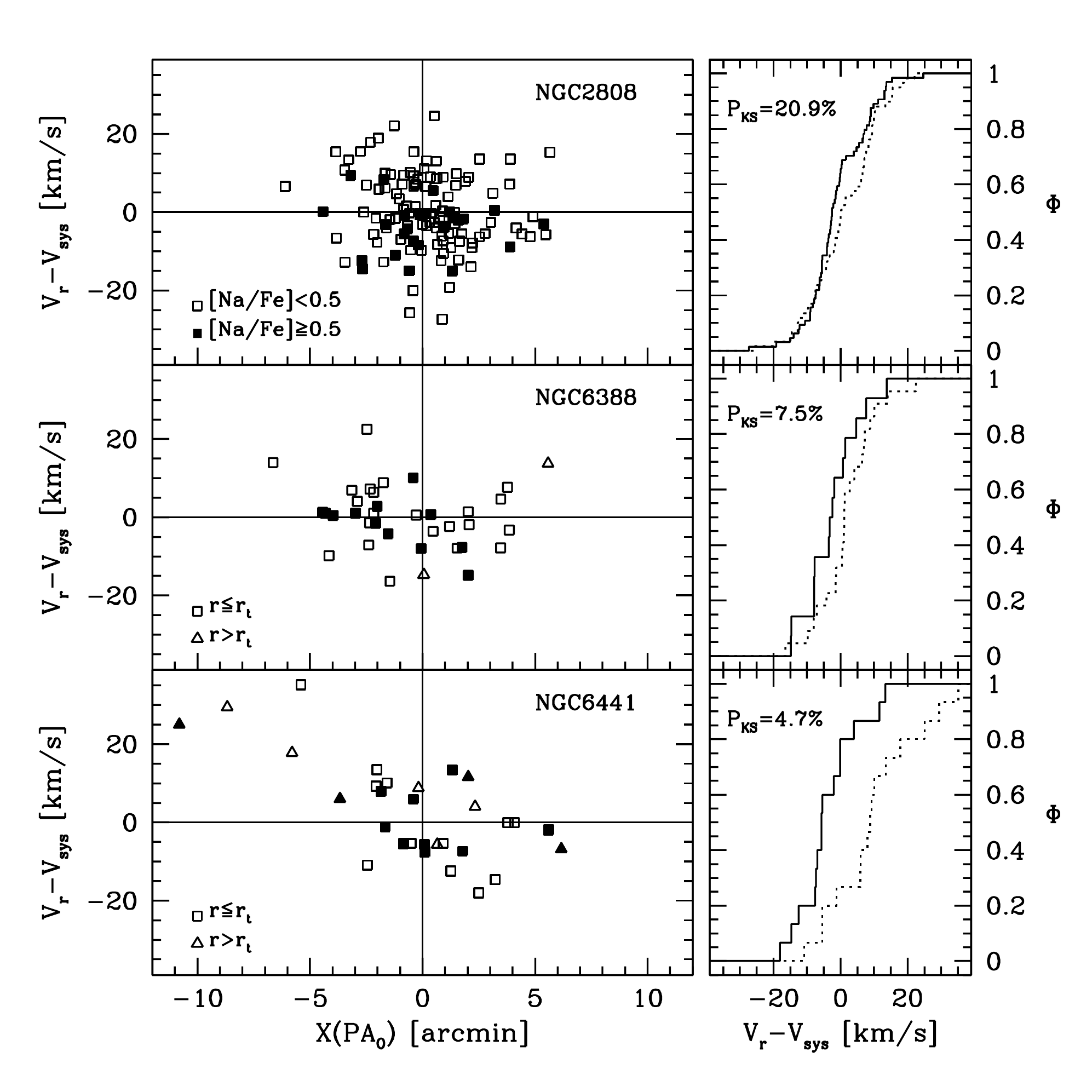}
     \caption{Rotation curves for NGC2808, NGC6388, and NGC6441. Filled and open symbols are adopted for Na-rich and Na-poor stars, respectively. {\em Left panels}: RV in the system of the cluster as function of distance from the center projected onto the axis perpendicular to the best-fit rotation axis found in Figs.~\ref{corot1} and \ref{corot2}. Stars inside and outside the tidal radius are plotted with different symbols for NGC6441 and NGC6388. All the stars in the NGC2808 sample lie within the tidal radius on the cluster. 
{\em Right panels:} comparison of the cumulative RV distributions of stars having $X(PA_0)>0.0$ (continuous lines) and $X(PA_0)<0.0$ (dotted lines). The probability that the two distributions are drawn from the same parent population (according to a KS test) is reported in each panel. We show rotation curves only for the two clusters having $P_{KS}<10$\%  plus NGC2808.}
        \label{curvrot}
    \end{figure}


In Fig.~\ref{curvrot} we show the rotation curve for the cases of NGC2808, NGC6388, and NGC6441, i.e. the distribution of RV as a function of the projection of the star position on the axis perpendicular to the axis of rotation [$X(PA_0)$]. In the righthand panels, the RV distribution of stars lying on opposite sides with respect to the rotation axis are compared. When there is no rotation, the two distributions should be indistinguishable, while with significant rotation a shift should be apparent. The degree to which the two distributions differ also depends on the ratio between rotation and velocity dispersion and on the actual shape of the rotation curve. In the considered cases the comparison is not fully conclusive, given the small samples under consideration. A KS test states that it is relatively unlikely (but clearly not impossible) that the observed patterns may emerge by chance from nonrotating systems. 
For NGC2808 we had the opportunity to fully confirm the result presented here with a much larger, still unpublished sample of RV estimates (more than 800 stars, Bragaglia \& Carretta, private communication).
The rotation pattern observed in NGC1851 has also been independently confirmed with an unpublished sample of $\sim 100$ HB stars observed within another program (Gratton \& Lucatello, private communication). 

Figure~\ref{curvrot} deserves a couple of further comments:

\begin{itemize}

\item In the upper lefthand panel of Fig.~\ref{curvrot} the subtle selection effect of the uneven spatial sampling in NGC2808  can be appreciated.  The effect of rotation (that favors a certain range of velocities for stars in a given wing of the rotation curve) coupled with 
low-number statistics may well be at the origin of the correlation between Na-abundance and velocity dispersion discussed above. In fact, in each wing the velocity dispersion is lower than in the two wings considered as a whole.

\item It is interesting to note that in NGC6441 stars within and outside the tidal radius seem to trace the same rotation pattern. The kind of solid body rotation extending to large distances from the cluster center that is hinted at in Fig.~\ref{curvrot} may be suggestive of a velocity gradient induced by tidal stripping \citep[see][discussion and references therein]{munoz,mic54,sollima}.

\end{itemize}

\begin{table*}
  \begin{center}
  \caption{Cluster parameters}\label{Tab_V}
  \begin{tabular}{lccccccccccccc}
    \hline
Clus.  &$\sigma_0$&$\epsilon_{\sigma}$&$A_{rot}$&$\epsilon_{A}$&[Fe/H]&IQR[Na/O]&$\frac{{B-R}}{{B+R+V}}$&$M_V$&  ell & log$\rho_0$ & $R_{GC}$ & $|Z|$ & Notes \\
       & \kms    & \kms	   &\kms	 &\kms     &      &      &	     &	    & 	&$L_{\sun}{\rm pc}^{-3}$&kpc&kpc&		\\
    \hline
       &     &	   &	 &     &      &      &	     &	    &	&&&&		\\
NGC104 &  9.6&  0.6&  4.4&  0.4& -0.76& 0.472& -0.99 &  -9.42&  0.09&  4.88 &	 7.4&  -3.1 &	L10b		\\
NGC288 &  2.7&  0.8&  0.5&  0.3& -1.32& 0.776&  0.98 &  -6.75&   ---&  1.78 &	12.0&  -8.9 &	L10b		\\
NGC1851& 10.4&  0.5&  1.6&  0.5& -1.16& 0.693& -0.32 &  -8.33&  0.05&  5.09 &	16.6&  -6.9 &	H96+t.w.$^a$	\\
NGC1904&  5.3&  0.4&  0.6&  0.5& -1.58& 0.759&  0.89 &  -7.86&  0.01&  4.08 &	18.8&  -6.3 &	H96+t.w.$^a$	\\
NGC2808& 13.4&  1.2&  3.3&  0.5& -1.18& 0.999& -0.49 &  -9.39&  0.12&  4.66 &	11.1&  -1.9 &	H96+t.w.	\\
NGC3201&  4.5&  0.5&  1.2&  0.3& -1.51& 0.634&  0.08 &  -7.45&  0.12&  2.71 &	 8.8&	0.7 &	C95 \\
NGC4590&  2.4&  0.9&  1.2&  0.4& -2.27& 0.372&  0.17 &  -7.37&  0.05&  2.57 &	10.2&	6.0 &	L10b		\\
NGC5024&  4.4&  0.9&  0.0&  0.5& -2.06&  --- &  0.81 &  -8.71&  0.01&  3.07 &	18.4&  17.6 &	L10b  \\
NGC5139& 19.0&  1.0&  6.0&  1.0& -1.64& 0.930&  ---  & -10.26&  0.17&  3.15 &	 6.4&	1.3 &	VV06+P07    \\
NGC5904&  7.5&  1.0&  2.6&  0.5& -1.33& 0.741&  0.31 &  -8.81&  0.14&  3.88 &	 6.2&	5.5 &	t.w.		\\
NGC6121&  3.9&  0.7&  1.8&  0.2& -1.18& 0.373& -0.06 &  -7.19&  0.00&  3.64 &	 5.9&	0.6 &	L10b		\\
NGC6171&  4.1&  0.3&  2.9&  1.0& -1.03& 0.522& -0.73 &  -7.12&  0.02&  3.08 &	 3.3&	2.5 &	H96+t.w.$^c$	\\
NGC6218&  4.7&  0.9&  0.3&  0.2& -1.33& 0.863&  0.97 &  -7.31&  0.04&  3.23 &	 4.5&	2.1 &	L10b		\\
NGC6254&  6.6&  0.8&  0.4&  0.5& -1.57& 0.565&  0.98 &  -7.48&  0.00&  3.54 &	 4.6&	1.7 &	H96+t.w.	\\
NGC6388& 18.9&  0.8&  3.9&  1.0& -0.45& 0.795& -0.65 &  -9.41&  0.01&  5.37 &	 3.1&  -1.2 &	H96+t.w.	\\
NGC6397&  4.5&  0.6&  0.2&  0.5& -1.99& 0.274&  0.98 &  -6.64&  0.07&  5.76 &	 6.0&  -0.5 &	MM91+t.w.      \\
NGC6441& 18.0&  0.2& 12.9&  2.0& -0.44& 0.660& -0.76 &  -9.63&  0.02&  5.26 &	 3.9&  -1.0 &	H96+t.w.	\\
NGC6656&  6.8&  0.6&  1.5&  0.4& -1.70&  --- &  0.91 &  -8.50&  0.14&  3.63 &	 4.9&  -0.4 &	L10b	  \\
NGC6715& 16.4$^c$&0.4&2.0&  0.5& -1.56& 1.169&  0.54 &  -9.98&  0.06&  4.69  &  18.9 & -6.5 &	    I09 + B08	    \\
NGC6752&  5.7&  0.7&  0.0&  0.0& -1.55& 0.772&  1.00 &  -7.73&  0.04&  5.04 &	 5.2&  -1.7 &	L10b		\\
NGC6809&  2.7&  0.5&  0.5&  0.2& -1.93& 0.725&  0.87 &  -7.57&  0.02&  2.22 &	 3.9&  -2.1 &	L10b		\\
NGC6838&  2.3&  0.2&  1.3&  0.5& -0.82& 0.257& -1.00 &  -5.61&  0.00&  2.83 &	 6.7&  -0.3 &	H96+t.w.	\\
NGC7078& 13.5&  0.9&  3.8&  0.5& -2.33& 0.501&  0.67 &  -9.19&  0.05&  5.05 &	10.4&  -4.8 &	H96+t.w.	\\
NGC7099&  5.0&  0.9&  0.0&  0.0& -2.33& 0.607&  0.89 &  -7.45&  0.01&  5.01 &	 7.1&  -5.9 &	L10b		\\
\hline
\end{tabular}
\tablefoot{The references in the last column indicate the source of the velocity dispersions and of
the rotation amplitudes (t.w.= this work; H96= 2010 version of the \citet{h96} catalog; B08=\citet{mic54}; I09=\citet{ibam54}; P07=\citet{panci07}; MM91=\citet{MM91}; C95=\citet{cote}; VV06=\citet{vandeven}; for the other acronyms see Tab.~\ref{Tab_sam}). 
The $A_{rot}$ values from L10b have been multiplied by 2 to report them on the same scale as ours.
The metallicity values are from \citet{scala}, except for NGC1851 \citep[from][]{n1851} and NGC6715=M54
\citep[from][]{m54all}. IQR[Na/O] values are taken from \citet{IQR} except for NGC6715 \citep[from][]{m54om} and 
NGC5139=$\omega$~Cen \citep[from][]{n1851}. $M_V$ and ell values are from H96.
$\frac{{B-R}}{{B+R+V}}$ values are taken from \citet{mack} except for NGC6388 and NGC6441; for these cluster we computed updated values
(properly accounting for the extended blue HB tail) from the data reported in Table~3 of \citet{busso}\\
$^a$ In agreement with the results by \citet{sca11}.
$^b$ In agreement with the results shown in \citet{sca04}.
$^c$ Excluding the inner cusp possibly associated with a central Black Hole \citep[see][for details]{ibam54}
} 
\end{center}
\end{table*}

We have to stress again that the results presented in this section may also suffer from biases associated with the dimension of the samples and with the radial distribution of sample stars. However, most of the stars at relatively large distances from the cluster centers may not be a strong concern for detecting rotation, since it is expected that the highest rotation amplitudes are reached relatively far from the center \citep{mmm97,sca11}. Moreover, the adopted technique is not expected to be particularly sensitive, since any rotation curve is strongly smoothed out in the simple average velocity of two halves of the clusters: as a result, the detected signals should be considered as quite robust and homogeneous.

\section{Summary and discussion} 
\label{summ}

   \begin{figure*}
   \centering
   \includegraphics[width=\columnwidth]{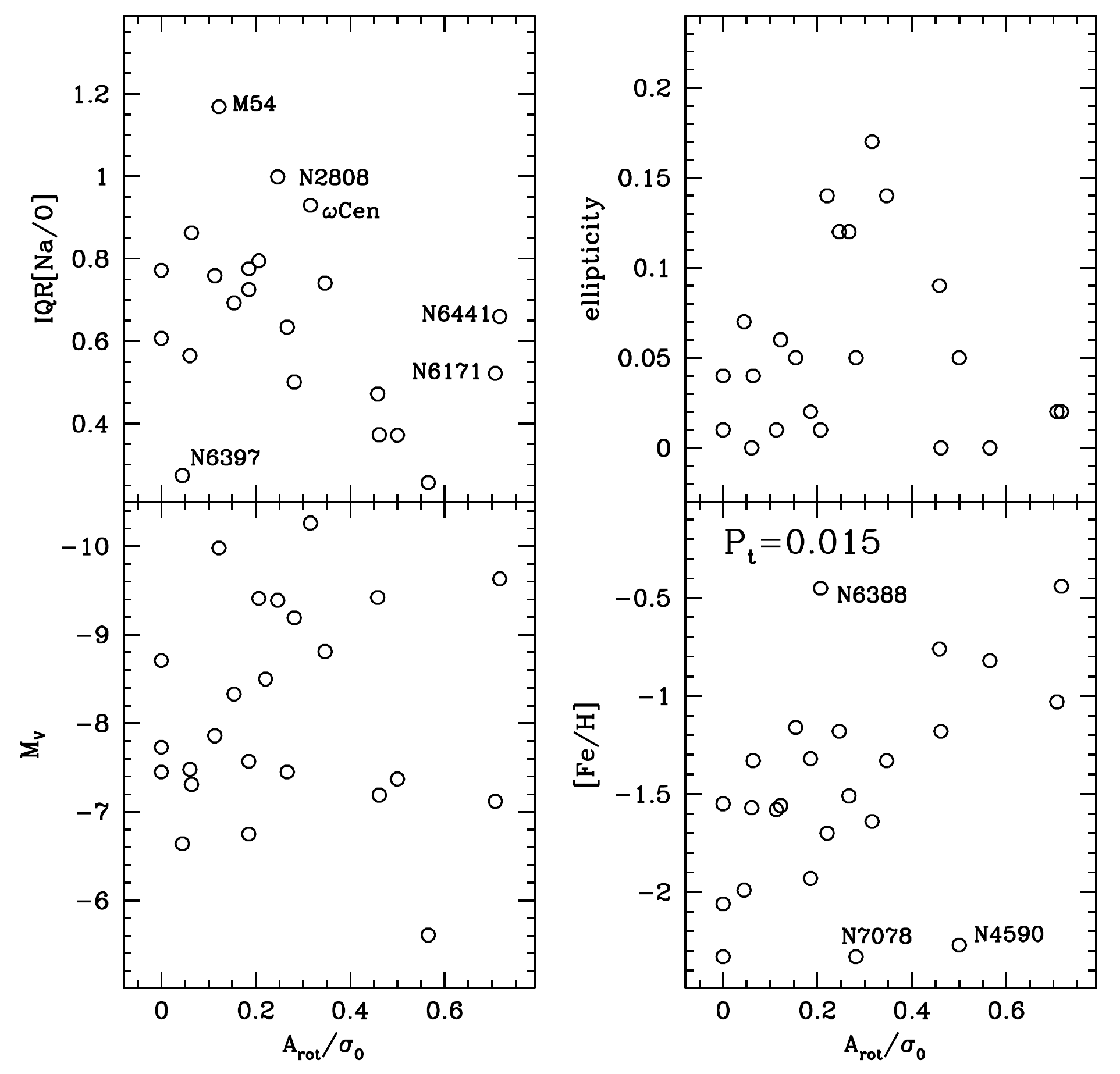}
   \includegraphics[width=\columnwidth]{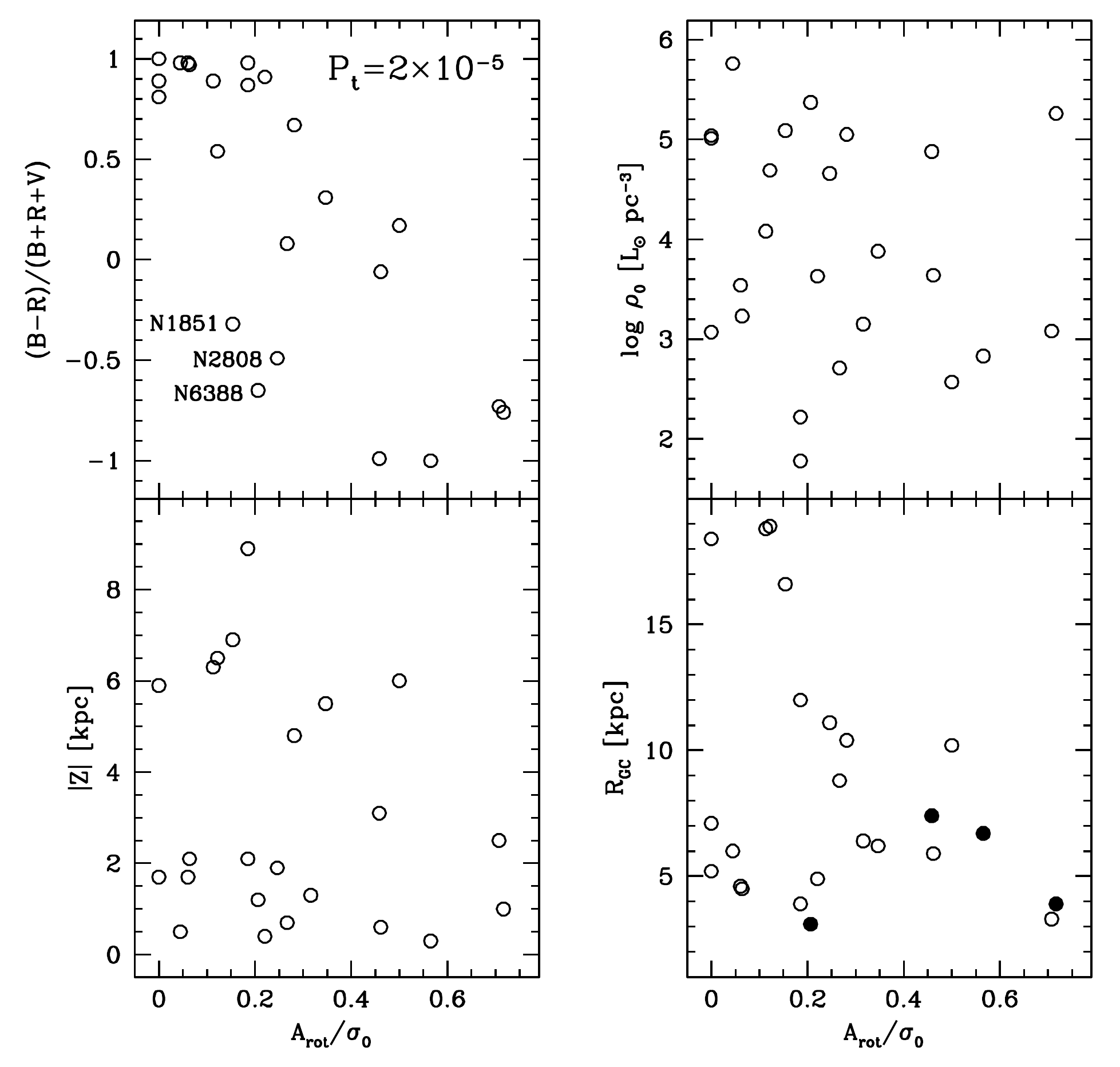}
     \caption{The ratio between the amplitude of the rotation $A_{rot}$ to the central velocity dispersion $\sigma_0$ is plotted versus various other parameters. The clusters that appear to lie at odd positions with respect to the main trends are labeled. In the rightmost panel of the lower row, clusters having [Fe/H]$\ge -1.0$ are plotted as filled circles.}
        \label{Vcorr}
    \end{figure*}


We used the radial velocity estimates obtained from a large spectroscopic survey (aimed at the study of the Na-O anticorrelation in GCs) to study the kinematics of the surveyed clusters, with the main aim of looking for correlations between kinematics and chemical abundance of elements involved in the anticorrelation issue. With all the limitations associated with the relatively small samples and the uneven radial distribution of target stars within each cluster, the data considered here allow  the exploration of this interesting issue for the first time. By internal and external comparisons we have demonstrated that the typical uncertainty on individual RV estimates is smaller than 1~\kms. 

In general, we did not find any evidence of correlation of Na abundance with velocity dispersion or rotation within each cluster. A possible decrease in velocity dispersion for [Na/Fe]$\ga 0.4$ for the peculiar clusters NGC6388 and NGC6441, in broad agreement with the expectations from the models by \citet{bekkimod}, clearly requires larger datasets to be confirmed or disconfirmed. A statistically significant 
decrease in velocity dispersion for [Na/Fe]$\ga 0.4$ for NGC2808 is found to likely come from an uneven spatial distribution of the target stars in this cluster. 

We investigated in slightly deeper detail the cases of the clusters whose kinematics were previously studied in the literature with smaller samples than ours. We confirm (or provide support for) the central velocity dispersion estimates reported in the literature for NGC6388, NGC2808, and NGC6254. We provide a revised estimate of the central velocity dispersion for NGC5904. 

Finally we looked for systemic rotation in the clusters that were lacking such analysis in the literature (or had been studied with different methods than that adopted here). We found that the considered clusters
cover a remarkable range of rotational amplitudes, from $\sim 0$~\kms to $\sim 13$~\kms. NGC5904, NGC6171, NGC2808, NGC6388, and NGC6441 show especially strong rotation patterns ($A_{rot}\ga 2.5$~\kms).

In Table~\ref{Tab_V} we list homogeneous $\sigma_0$ and $A_{rot}$ estimates collected from this work and other sources in the literature (mainly L10b), together with other relevant cluster parameters from various sources (mainly from H96). We added to our original sample $\omega~Cen$ (NGC5139), M54 (NGC6715), M22 (NGC6656), and M53 (NGC5024) for which we had all (or most of) the relevant data in the same scales adopted here from other sources \citep[L10b; B08;][]{m54om,panci07}. 
It is interesting to check that the rotation properties of the clusters do correlate with some other characteristics, as this is the largest collection of rotational amplitudes for Galactic GCs currently available \citep[see][for a summary of the data available {\em circa} one decade ago]{MH97}. Before going on with the analysis it may be worth recalling that projection effects should be at work, and the listed $A_{rot}$ are only estimates of the amplitude of the {\em projected} rotation pattern, thus lower limits of the actual rotation amplitude. {\em However, in Appendix~\ref{proxy} we show that $A_{rot}$ is a reasonable proxy for the true amplitude, in a statistical sense \citep[see also][]{chandra}.}

As the most informative parameter, we use the ratio between the rotation velocity $V_{rot}$, here represented by $A_{rot}$, and the central velocity dispersion $\sigma_0$. In Fig.~\ref{Vcorr}
we show scatter plots of $A_{rot}/\sigma_0$ vs. the inter quartile range (IQR), a parameter defined by \citet{eugesolo} to measure the extension of the Na-O distribution in clusters, the cluster ellipticity, the absolute integrated V magnitude ($M_V$), the iron abundance (metallicity, [Fe/H]), the classical HB morphology parameter $\frac{{B-R}}{{B+R+V}}$\footnote{Where B,R, and V are the number of HB stars lying to the blue, to the red, and within the RR Lyrae instability strip, respectively.} \citep{lee}, the logarithm of the central luminosity density (${\rm log}\rho_0$), the absolute value of the distance from the Galactic plane ($|Z|$), and the distance from the Galactic center ($R_{GC}$). 

The ratio $A_{rot}/\sigma_0$ does not display any clear correlation with $M_V$, ellipticity, log$\rho_0$, $|Z|$, and $R_{GC}$. The lack of a connection with ellipticity may appear somehow surprising. However, it must be recalled that the two quantities may refer to different regions of the clusters, since it is hard to reliably estimate ellipticity at large distances from the cluster core. On the other hand, a hint of a trend with IQR is perceived. In general clusters with higher $A_{rot}/\sigma_0$ ratios have smaller IQR. Four of the five clusters lying on a nearly parallel sequence above the bulk of the other objects (at larger IQR for a given $A_{rot}/\sigma_0$ value) are among the most peculiar clusters, two of them also displaying a spread in the iron abundance. According to a two-tailed Student's test, 
the probability that a Spearman rank correlation coefficient equal or larger (in absolute value) than the observed one ($s=-0.353$) is produced by chance from uncorrelated quantities is $P_t=$12.6 per cent, so the correlation cannot be considered as statistically significant \citep[][see Table~\ref{Tab_Asig} for $s$ and $P_t$ for all the considered parameters]{numrec}.
For this reason we refrain from any interpretation of this plot, which can be considered as a possible starting point for further analysis.

An interesting trend of $A_{rot}/\sigma_0$ is with the cluster metallicity: clusters with higher metallicity have greater fractions of ordered motion with respect to pressure support. In spite of the three outliers, the correlation is found to be remarkably significant by the same kind of Student test as used above, $P_t=$1.5 per cent. Since larger metal content in a gas implies higher efficiency in energy dissipation by atomic transitions, the observed correlation, if confirmed, may hint at a significant role of dissipation in the process of cluster formation \citep[see][]{bekkirot,bekkimod}.

The most significant correlation of $A_{rot}/\sigma_0$, however, is with the HB morphology ($\frac{{B-R}}{{B+R+V}}$): the relevance of ordered motions with respect to pressure is stronger for clusters with redder HBs. The probability that the observed correlation arose by chance is negligible ($P_t=2\times 10^{-5}$). It is well known that $\frac{{B-R}}{{B+R+V}}$ and [Fe/H] are correlated \citep[since metallicity is the {\em first} parameter determining the HB morphology, see][and references therein]{lee,f93,2nd3rd}, so it is likely that the correlation of one of the two parameters with $A_{rot}/\sigma_0$ is a secondary effect of the correlation with the other. We cannot provide a firm identification of the primary correlation; however, we note that for our sample the correlation of $A_{rot}/\sigma_0$ with $\frac{{B-R}}{{B+R+V}}$ (s=-0.769) is much stronger than both that of $A_{rot}/\sigma_0$ with [Fe/H] (s=0.492) and of $\frac{{B-R}}{{B+R+V}}$ with [Fe/H] (s=0.661). As a result, the connection seems more direct with the HB morphology than with the metallicity\footnote{A possible connection between cluster systemic rotation and HB morphology was already suggested more than 25 years ago by \citet{BCF85}, inspired by a previous suggestion by \citet{norris_rot}. However, the chain of processes envisaged by these authors was expected to produce {\em bluer} HB morphologies for greater degrees of systemic rotation, i.e. a correlation in the opposite sense with respect to what is found here.}. It is interesting to note that $\frac{{B-R}}{{B+R+V}}$ shows some degree of correlation 
($|s| \sim 0.3$; not shown here) with other HB morphology parameters considered in \citet[][like e.g., ${\rm log} T_{eff,HB}^{max}$]{IQR} and/or estimates of Helium abundance \citep[e.g. $Y_{max}$, ][]{2nd3rd}, in the sense that clusters with larger ${\rm log} T_{eff,HB}^{max}$ and $Y_{max}$
have lower $A_{rot}/\sigma_0$. These parameters, in turn, positively correlate with IQR that, as seen above, displays an anticorrelation with $A_{rot}/\sigma_0$. In summary, the overall observational scenario strongly suggests a significant link between the systemic rotation of the clusters and the processes that lead to the formation and evolution of the different generations of cluster stars, whose effects are likely imprinted (and most evident) in the current HB morphology \citep[][and references therein]{f93,2808,2nd3rd}.

It should also be noted that the orbits of the most metal-rich (and HB-red) clusters are likely confined within the Galactic bulge, so they may be subject to stronger tides, possibly producing velocity gradients that can mimic rotation \citep[see e.g.,][]{munoz}. However, in this case one would expect to see a better correlation between cluster position and $A_{rot}$ than with
$A_{rot}/\sigma_0$, since the amplitude of the gradient should mainly depend on the orbital parameters. The Spearman rank correlation coefficients listed in Table~\ref{Tab_Acorr} show that this is not the case; the correlation is poor also between
$A_{rot}$ and the orbital parameters of the clusters \citep[$R_{per}$ and $Z_{max}$ from][see Table~\ref{Tab_Acorr}]{dana99,dana07}. 

\begin{figure}
   \centering
   \includegraphics[width=\columnwidth]{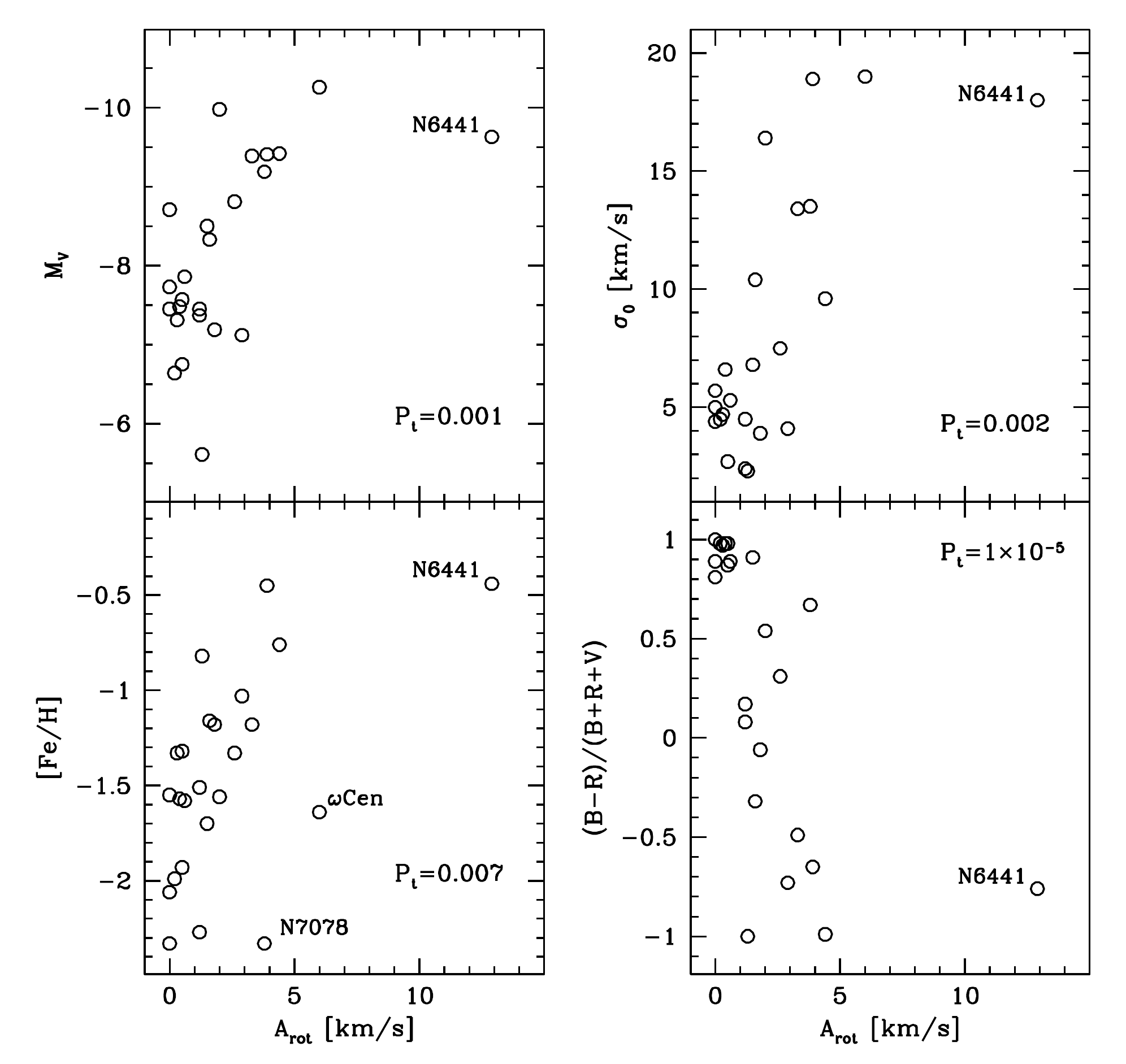}
     \caption{$A_{rot}$ vs. the four parameters displaying the strongest correlation with it (Tab.~\ref{Tab_Acorr}). 
     In all cases, removing NGC6441 from the sample has a negligible effect on
     the final value of $P_t$.}
        \label{Acorr}
    \end{figure}

On the other hand, $A_{rot}$ displays significant correlations with $\frac{{B-R}}{{B+R+V}}$, $M_V$, $\sigma_0$, and [Fe/H] ($P_t<$~1.0 percent, in all cases; see Fig.~\ref{Acorr}). 
Also in this case, the strongest and most significant correlation is, by far, with HB morphology,  therefore $A_{rot}$ seems strictly tied with {\em intrinsic} cluster parameters, not with their orbit or environment. Note that $M_V$ and $\sigma$ are also correlated between each other, both approximately tracing the cluster ranking in mass, but do not appear to correlate with [Fe/H] and/or $\frac{{B-R}}{{B+R+V}}$. 
The only outlier of the correlations of $A_{rot}$ with $M_V$ and $\sigma_0$ is NGC6441, which has a much larger $A_{rot}$ for its $M_V$ or $\sigma_0$ than do the other clusters of the sample. This may support the idea of a tidal origin for the strong velocity gradient observed {\em in this specific cluster}, as already suggested in Sect.~\ref{rot}. For the other clusters rotation is more likely of internal origin, possibly linked with the phase of cluster formation, with tides as a possible source of noise in the correlations. Removing NGC6441 from the sample has a negligible effect on the Spearman coefficients and on the statistical significance of the correlations, as measured by $P_t$.
We found two linear combinations of parameters displaying correlations with $A_{rot}$ that are significantly stronger than all those with single parameters, $LC1=\frac{{B-R}}{{B+R+V}}+0.47M_V$ and $LC2=M_V-1.73$[Fe/H]. The significance of both correlations is very high ($P_t\le 10^{-5}$). It is interesting to note that $M_V$ (likely as a proxy of the total cluster mass) has been found to play a role in bi-variate correlations, both with some HB morphology parameters \citep{f93,recio} and with parameters related to the anticorrelation phenomenon \citep[Pap-VII,][]{p8uve}.

As typical of most exploratory studies, our analysis does not provide firm conclusions, but calls for further investigations and perhaps opens some new windows on the research on globular clusters. It is clear that a follow up of the possible chemistry-kinematics connection and of the rotation patterns detected here is necessary for NGC2808 and, above all, for NGC6388 and NGC6441. The possible correlation between $A_{rot}/\sigma_0$  and IQR also deserves further study. This requires a considerable observational effort, e.g, to double the number of GCs with relatively large sample of Na abundances and RV estimates. The correlations between $A_{rot}/\sigma_0$, $\frac{{B-R}}{{B+R+V}}$  and [Fe/H] should be easier to verify and/or extend, since only large RV samples are required, with no need of detailed chemical abundance analysis of individual stars. The same is true for the correlations between $A_{rot}$ and $\frac{{B-R}}{{B+R+V}}$, $M_V$, $\sigma_0$, and [Fe/H]. These, together with those of $A_{rot}/\sigma_0$ with HB morphology and metallicity, are especially interesting since they are quite significant, from a statistical point of view, and far from trivial, since they indicate that rotation may be intimately linked with the internal physics of the clusters, and, possibly, with their origin.

\begin{table}
  \begin{center}
  \caption{Spearman rank correlation coefficients for $A_{rot}/\sigma_0$}\label{Tab_Asig}
  \begin{tabular}{lcccc}
    \hline
Parameter  & Spearman $s^a$ & $|t|^b$ & $N_{clus}^c$ & $P_t^d$\\ 
    \hline
       &       &	   &	 & \\
(B-R)/(B+R+V)  &-0.769   &5.526  &23  &  2$\times10^{-5}$\\
${\rm [Fe/H]}$ & 0.492   &2.652  &24  &  0.015 \\
IQR            &-0.336   &1.595  &22  &  0.126 \\  
Z              & 0.271   &1.322  &24  &  0.200 \\
$R_{GC}$       &-0.211   &1.014  &24  &  0.321 \\
log$\rho_0$    &-0.196   &0.936  &24  &  0.359 \\
$Z_{max}^{*}$  &-0.188   &0.835  &21  &  0.414 \\
$R_{per}^{*}$  &-0.187   &0.829  &21  &  0.417 \\
ell            & 0.145   &0.673  &23  &  0.508 \\
$M_V$          &-0.076   &0.360  &24  &  0.722 \\
$\sigma_0$     &-0.010   &0.049  &24  &  0.961 \\
\hline
\end{tabular}
\tablefoot{
$^a$ Spearman rank correlation coefficient of $A_{rot}/\sigma_0$ vs. the considered parameter.\\
$^b$ Absolute value of the associated $t$ variable as defined in \citet{numrec}.\\
$^c$ Number of clusters involved in the correlation. $t$ is approximately distributed as 
a Student distribution with $N_{clus}-2$ degrees of freedom \citep{numrec}.\\ 
$^d$ Probability to obtain a value of $s$ higher than or equal to observed, in absolute 
value [i.e. $P(|s|\ge |s_{obs}|)$] from two independent variables 
(i.e. in absence of correlation), from a Student test.\\
$^{*}$ Peri-galactic distance ($R_{per}$) and maximum distance above the Galactic plane ($Z_{max}$) are taken from
\citet{dana99,dana07}, as in \citet{IQR}.
} 
\end{center}
\end{table}

\begin{table}
  \begin{center}
  \caption{Spearman rank correlation coefficients for $A_{rot}$}\label{Tab_Acorr}
  \begin{tabular}{lcccc}
    \hline
Parameter  & Spearman $s^a$ & $|t|^b$ & $N_{clus}^c$ & $P_t^d$\\ 
    \hline
       &      &	   &	 & \\
(B-R)/(B+R+V) &-0.775   &5.614  &23  &  1$\times10^{-5}$\\
$M_V$         &-0.611   &3.623  &24  &  0.001 \\
$\sigma_0$    & 0.604   &3.558  &24  &  0.002 \\
${\rm [Fe/H]}$& 0.531   &2.941  &24  &  0.007 \\
ell           & 0.358   &1.755  &23  &  0.092 \\
log$\rho_0$   & 0.245   &1.186  &24  &  0.248 \\
$R_{per}^{*}$ &-0.244   &1.097  &21  &  0.286 \\
IQR           & 0.117   &0.526  &22  &  0.605 \\  
$R_{GC}$      &-0.109   &0.512  &24  &  0.614 \\
$Z_{max}^{*}$ &-0.106   &0.467  &21  &  0.646 \\
Z             &-0.048   &0.224  &24  &  0.825 \\
\hline
LC1$^a$          &-0.888   &8.865  &23  & 2$\times10^{-8}$ \\
LC2$^b$          &-0.811   &6.509  &24  & 1$\times10^{-6}$ \\
\hline
\end{tabular}
\tablefoot{
The meaning of the symbols is the same as in Tab.~\ref{Tab_Asig}.\\
$^a$ LC1=$(B-R)/(B+R+V)+0.47M_V$, is the linear combination of the pair of listed parameters 
displaying the strongest correlation with $A_{rot}$.\\
$^b$ LC2=$M_V-1.73$[Fe/H], displaying a correlation with $A_{rot}$ nearly as strong as LC1.\\
} 
\end{center}
\end{table}

\begin{acknowledgements}
This paper is dedicated to the memory of Robert T. Rood, whose contribution to the understanding of the Horizontal Branch phase of the evolution of low mass stars cannot be overstated.
We acknowledge the financial
support to this research by INAF through the PRIN-INAF 2009 grant CRA 1.06.12.10 (PI: R. Gratton).

\end{acknowledgements}

\bibliographystyle{apj}


\appendix
\section{The projected rotation amplitude as a tracer of the true rotation amplitude}
\label{proxy}

To explore quantitatively the effect of projection on the distribution
of observed rotation amplitudes of a population of GCs (or any other system whose inclination w.r.t. the plane of the sky $i$ is unknown and unconstrained), we performed the following simulation.
In a cluster with a {\em true} rotation amplitude $A_{true}$, we detect an {\em observed} rotation amplitude $A_{obs}=A_{true}{\rm sin}(i)$, where $0\degr\le i\le 90\degr$ ($i=0\degr$ corresponds to a cluster seen pole-on, $i=90\degr$ corresponds to a cluster seen ``edge-on'', i.e. having $A_{obs}=A_{true}$). 

   \begin{figure}
   \centering
   \includegraphics[width=\columnwidth]{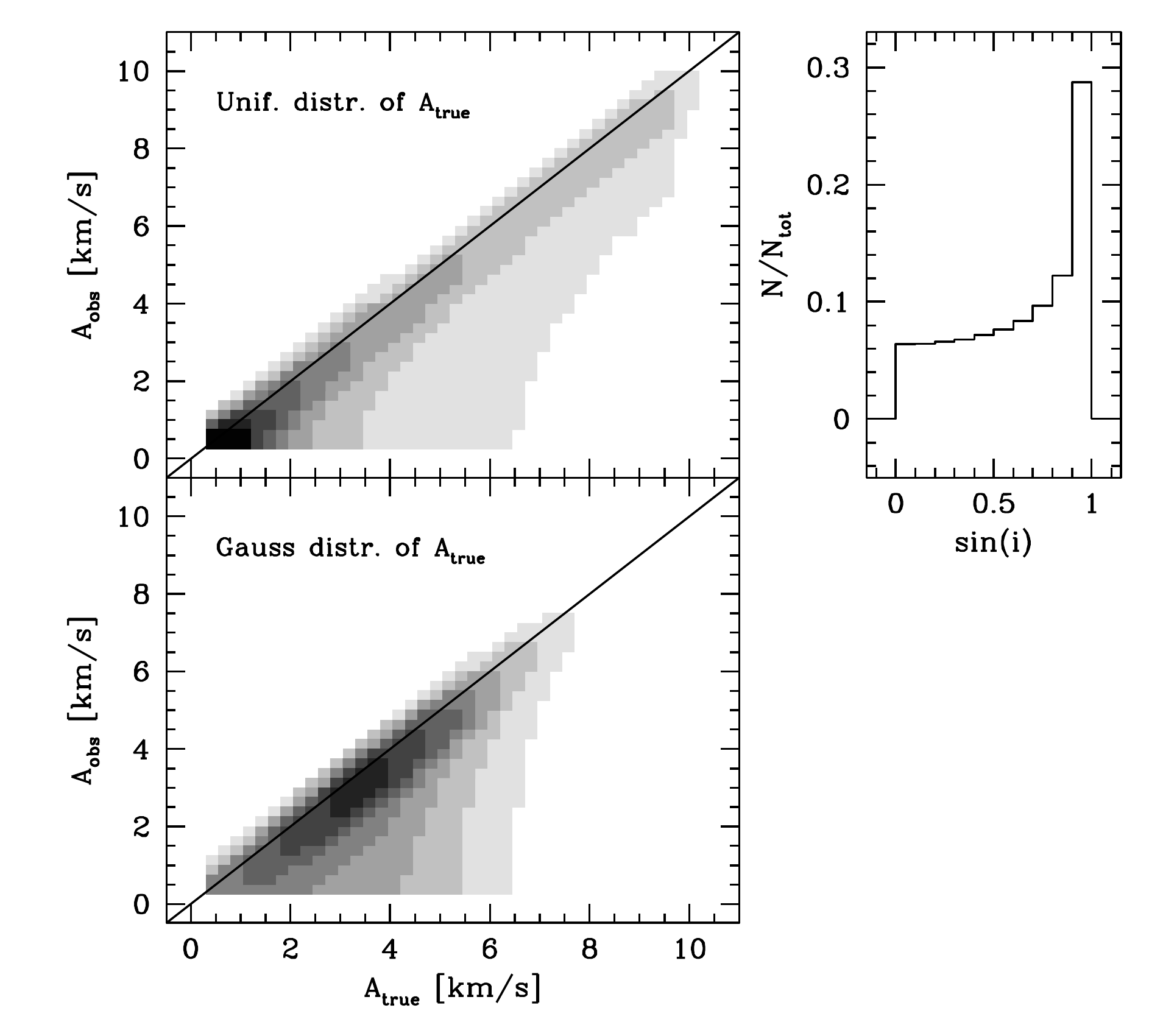}
     \caption{{\em Left panels:} probability distribution of observing a rotation amplitude 
     ($A_{obs}$) given a true value of the amplitude ($A_{true}$) and a uniform distribution of inclination angles ($i$).
Different tones of gray indicate different probability levels, from 8\% (darkest gray) to 1\% (lightest gray), in steps of 1\%; white regions corresponds to probability below 1\%.
The continuous line is the $A_{obs}=A_{true}$ relation. In the upper panel 
$A_{true}$ values are drawn from a uniform distribution, in the lower panel they are drawn from a Gaussian distribution.
The probability is estimated in square bins with sides of 1.0~\kms~ spaced by 0.25~\kms~ in both directions of the plane.
{\em Right Panel:} distribution of sin($i$) for a uniform distribution of $i$.
}
        \label{traceA}
    \end{figure}


We simulated a population of $10^6$ clusters whose inclination is drawn from a uniform distribution, since there is no preferred value for $i$. We explored two different distributions of $A_{true}$: a uniform distribution in the range 0.0~\kms $\le A_{true}\le$ 10.0~\kms, and a Gaussian distribution
with mean $\langle A_{true}\rangle=$4.0~\kms and $\sigma=$2.0~\kms. In the lefthand panels of Fig.~\ref{traceA} we plot the density distributions obtained in the two cases in the plane $A_{true}$ vs. $A_{obs}$. It is quite clear that the regions close to the $A_{obs}=A_{true}$ relation display the highest probability density. Independently of the assumed distribution of $A_{true}$, for a uniform distribution of inclinations, observed rotation amplitudes are thus more likely close to the true value.

This is due to the simple fact that the sinus function maps a uniform distribution of angles $i$ between $0\degr$ and $90\degr$ into a distribution of sin($i$) that is strongly peaked toward 
sin($i$)=1.0, as shown in the upper righthand panel of Fig.~\ref{traceA}. In particular sin($i$)$\ge 0.5$ in $\simeq$66\% of the cases, sin($i$)$\ge 0.7$ in $\simeq$50\% of the cases, and sin($i$)$\ge 0.9$ in $\simeq$28\% of the cases, while  sin($i$)$< 0.2$ in just $\simeq$12\% of the cases. We can therefore conclude that $A_{obs}$ is a reasonable proxy for $A_{true}$ in a statistical sense. This supports the idea that the observed correlations between $A_{rot}$ or $A_{rot}/\sigma_0$ and, e.g., $M_V$, [Fe/H] or $\frac{{B-R}}{{B+R+V}}$, may trace physical correlations between these parameters and the {\em true} rotational amplitude.


\end{document}